\documentclass [11pt, letterpaper]{article}
\pdfoutput=1 
\usepackage{fullpage}
\usepackage{amsmath}
\usepackage{amssymb}
\usepackage{graphicx}
\usepackage{braket}
\usepackage{mathrsfs}
\usepackage{wrapfig}
\usepackage{boxedminipage}
\usepackage{epsfig}
\usepackage{setspace}
\usepackage{subfigure}
\usepackage{float}
\usepackage{hyperref}
\usepackage{enumerate}
\usepackage{cite}
\usepackage{bbold}
\usepackage[font=footnotesize,labelfont=rm]{caption}

\begin{document}

\begin{flushright}
{\tt arXiv:1907\!.05883}
\end{flushright}

{\flushleft\vskip-1.35cm\vbox{\includegraphics[width=1.25in]{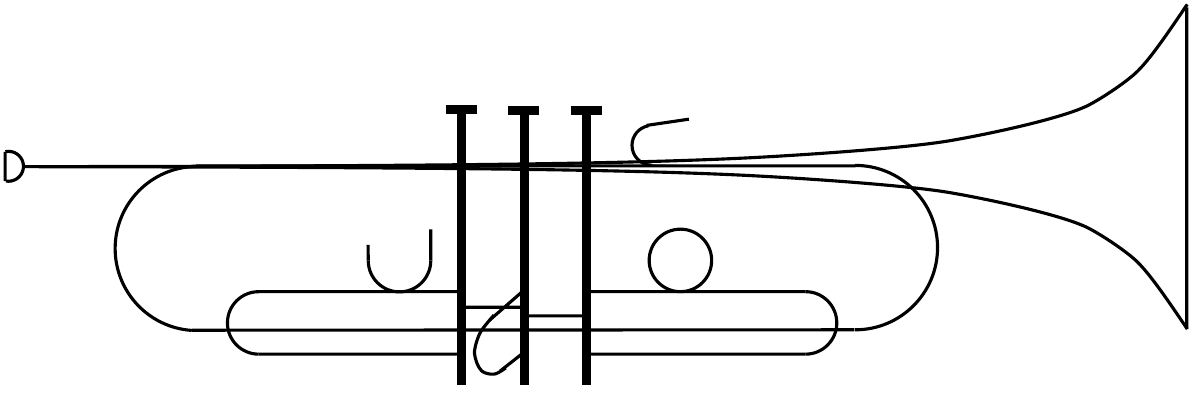}}}

\bigskip
\bigskip

\bigskip
\bigskip
\bigskip
\bigskip

\begin{center} 

{\Large\bf   de Sitter Black Holes, Schottky Peaks}

\bigskip \bigskip 

{\Large\bf and}

\bigskip \bigskip 
{\Large\bf Continuous  Heat Engines}

\end{center}

\bigskip \bigskip \bigskip \bigskip

\centerline{\bf Clifford V. Johnson}

\bigskip
\bigskip
\bigskip

\centerline{\it Department of Physics and Astronomy }
\centerline{\it University of
Southern California}
\centerline{\it Los Angeles, CA 90089-0484, U.S.A.}

\bigskip

\centerline{\small {\tt johnson1}  [at] usc [dot] edu}

\bigskip
\bigskip


\begin{abstract} 
\noindent 
Recent work has uncovered Schottky--like peaks in the temperature dependence of  key specific heats of certain black hole thermodynamic systems. They  signal a finite window of available energy states for the underlying microscopic degrees of freedom. This paper reports on new families of  peaks, found for the Kerr and Reissner--Nordstr\"om black holes in a spacetime with positive cosmological constant. It is known that a system with a highest energy, when coupled to  two distinct heat baths, can naturally generate a thermodynamic instability,  population inversion, a channel for work output.  It is noted that these features are all present for de Sitter black holes. It is shown that there are trajectories in parameter space where they  behave as generalized masers,  operating  as  continuous heat engines, doing work by shedding angular momentum. It is suggested that  bounds on efficiency due to the second law of thermodynamics  for  general de Sitter black hole solutions could provide powerful consistency checks.
\end{abstract}

\pagenumbering{gobble}

\newpage 

\pagenumbering{arabic}

\baselineskip=18pt 
\setcounter{footnote}{0}

\section{Introduction}
\label{sec:introduction}

An examination of the behaviour of  certain thermodynamic quantities can provide a wealth of useful information about the underlying microscopic quantum mechanical degrees of freedom, sometimes providing key clues as to the formulation of the underlying model. This has a long history, starting with early models of gases provided by Einstein~\cite{Einstein1907} (refined by Debye~\cite{Debye1912}),  that derive the temperature dependence of the specific heat at constant volume,  $C_V(T)$  starting  from simple statistical mechanical considerations of the basic degrees of freedom (for a review see {\it e.g.,} refs.~\cite{mandl1995statistical,rosser1982introduction,kittel1996introduction}). Since~$C_V(T)$ is a macroscopic (and easily measurable) quantity, it provides a readily accessible window on the quantum underpinnings of matter.

In the search for a complete understanding of the quantum theory underlying  gravity, it is natural to regard black hole thermodynamics\footnote{Meaning both traditional~\cite{Bekenstein:1973ur,Bekenstein:1974ax,Hawking:1974sw,Hawking:1976de} and the  extended flavour developed in refs.~\cite{Kastor:2009wy,Wang:2006eb,Sekiwa:2006qj,LarranagaRubio:2007ut}. (See also refs.~\cite{Henneaux:1984ji,Henneaux:1989zc,Teitelboim:1985dp}.)} as the macroscopic ``measurement''. It is therefore wise to seek  quantities that might contain compelling clues about the nature and organization of the microscopic degrees of freedom. In that spirit, it was noticed recently that examining the temperature dependence of the specific heat at constant volume $C_V(T)$ for a large class of black holes (with negative cosmological constant $\Lambda$) revealed Schottky--like peaks~\cite{Johnson:2019vqf}. Just as jumps in $C_V(T)$ (familiar for diatomic gases, for example) signal the onset of new degrees of freedom available at higher energies (such as rotational modes), a fall--off of $C_V(T)$ is a clear signal of the absence or sparsity of degrees of freedom excited at higher energies. It was therefore suggested that, perhaps surprisingly, those black holes (in extended thermodynamics, so with a dynamical pressure~$p=-\Lambda/8\pi$) have a small and finite window of energies available for the degrees of freedom\footnote{This tunably small sector was used as the basis for making contact between black hole heat engines and quantum heat engines in ref.~\cite{Johnson:2019olt}.}. Studying $C_V(T)$ has also proven useful as a diagnostic for certain other black hole spacetimes, where it 
turned out to be negative, signalling a new kind of thermodynamic instability~\cite{Johnson:2019mdp}. It is important to note that the standard AdS/CFT application of such black holes are at fixed (negative) cosmological constant and hence fixed pressure. For such cases, the usual specific heat is equivalent to $C_p(T)$, and it  displays no such energy restriction for the degrees of freedom.

One of the main purposes of this paper is to point out that Schottky--like peaks are not just the domain of negative cosmological constant, or of extended thermodynamics, topics which may not be to the tastes of some researchers. They are a core feature of standard black hole thermodynamics when one considers positive cosmological constant. The cases of Kerr--de Sitter (Kerr--dS) and Reissner--Nordstr\"om--de Sitter (RN--dS) are reported here, and some lessons about the underlying microscopic theory are suggested. These systems contain, as a special case, the zero spin ($a$) and zero charge ($Q$) example of Schwarzschild--de Sitter, which also has a peak. While this project was being completed, the authors of ref.~\cite{Dinsmore:2019elr} reported the Schottky--like peak for that case, in  independent work. They also offered ideas about the nature of the underlying theory, examining whether  a model of a constrained spin system can capture features of the black hole\footnote{Using a spin system to model aspects of  black hole physics  was also briefly considered in ref.~\cite{Johnson:2019olt}, for the constant~$V$ sector.}, with interesting but mixed success. 

The interpretation of the thermodynamics that is presented in this paper takes a  somewhat different route. As suggested in further considerations of the Schottky--like peaks in ref.~\cite{Johnson:2019olt}, such structures can arise from simply truncating  a quantum system, rather than being a result of a small discrete system such as one provided by spins. The macroscopic phenomenon of the peak is similar in each case. This will be reviewed in section~\ref{sec:schottky-matters}, for completeness, using the case of a simple truncation of the harmonic oscillator as an example. The suggestion here is that de Sitter  black holes may be best modelled as a truncated quantum system  rather than as finite spin systems, allowing for the possibility that the quantum system otherwise has enough structure to reproduce much of the other known physics of de Sitter black holes, including the intermediate--distance physics that they share with their flat spacetime cousins\footnote{It is possible that the two approaches can meet in the middle, with sufficient complication added to a spin system so as to be equivalent to a truncated quantum system.}.

A core idea also presented in this paper concerns the nature of the two natural temperatures of these spacetimes. There is a temperature associated to the black hole horizon and one associated to the cosmological horizon. The system is not in thermodynamic equilibrium, which provides some tension with the fact that each horizon is  associated with a temperature, which is after all a quantity best understood in equilibrium thermodynamics. The idea here is that the energy cutoff actually makes this all extremely natural. It is known from other important quantum systems that two natural instabilities are present if there is a certain kind of coupling to two heat baths at different temperatures, together with a channel for decay that allows for some non--thermal ejection of energy.  The obvious (mild) instability is simply the flow of heat between heat baths, but the more interesting instability is the phenomenon of population inversion, familiar from lasers and masers. It is not possible without an effective highest energy state. Once population inversion is achieved, the system tries to restore  equilibrium not just with heat flow but by doing work, for as long as the situation remains. This is what is known as a continuum heat engine, and the maser was given such a (quantum) thermodynamic interpretation long ago by Scovil and Schulz--Du Bois~\cite{PhysRevLett.2.262}.

What will be shown here is that all of these mechanisms are available for de Sitter black holes. There is a cap on the energy, as signalled by the Schottky-like peaks, and so more than one heat bath will be natural. The Schwarzschild--de Sitter case is too simple to show population inversion (there is no channel for doing work), and so the only instability available there is to flow heat between the two heat baths. Kerr--dS and Reissner--Nordstr\"om--dS  on the other hand, are much richer, and they can proceed by a different, more interesting path. They have a channel for doing work, and in fact it is by reducing angular momentum (or charge). The classic Penrose process~\cite{Penrose:1969pc,Penrose:1971uk} (and the analogous process for charged black holes~\cite{Denardo:1973pyo}) is one way they can do this, resulting in black hole mass loss, but for continuous engines processes which keep the mass (internal energy) fixed will be the focus\footnote{In  earlier versions of this manuscript, a mis--identification of the sign of $Q_C$ meant that the first law was written incorrectly, and as a result the heat engine trajectories were mis--identified.}. 
 
As will become clear, constant mass paths in the parameter space of the dS black hole  system  have all the elements of a  continuous heat engine (an amusing complement to the reciprocating heat engine construction presented for AdS black holes in extended thermodynamics in ref.~\cite{Johnson:2014yja}).  The next idea is crucial: Assuming that the thermodynamics of the de Sitter black hole system is complete, and is to be taken seriously, {\it the implications of the second law of thermodynamics must be considered}. How the system explores parameter space should take it into account. The efficiency of conversion of heat into work is bounded by the second law. While the cases studied here (Kerr--dS and Reissner--Nordstr\"om--dS) will satisfy the second law in a somewhat trivial way, a general bound (in two forms) is proposed in this paper (amounting to a Carnot efficiency for the continuous heat engine), and more general de Sitter black hole solutions (possibly coupled to other fields) should satisfy it.
The full ramifications of this suggestion are left for future work. 

The outline of  the paper is as follows. In section~\ref{sec:Kerr-RN-dS},  the Schottky--like peaks for  Kerr--dS and  Reissner--Nordstr\"om--dS   are  presented. There are several new features not available for the Schwarzschild--dS case, and they are highlighted. In section~\ref{sec:schottky-matters}, there is a  discussion about the interpretation of the Schottky--like peaks, essentially recalling what was  stated in refs.~\cite{Johnson:2019vqf,Johnson:2019olt}. Section~\ref{sec:two-baths} presents some interpretation of the thermodynamics, with the continuous heat engine construction in subsection~\ref{sec:heat-engine}. The  bound is presented (in a strong and a weak form) and discussed in subsection~\ref{sec:bound}. There is a concluding discussion in section~\ref{sec:discussion}.
 
 It is worth making  a final general remark at this juncture.
Black holes are rather robust and generic. They dynamically intrude on any discussion about the microscopic degrees of freedom of short distance physics in gravity, and so the lessons they suggest should be taken seriously. This is what led  't Hooft  to the holographic principle of quantum gravity~\cite{tHooft:1993dmi}, which underpins powerful frameworks like AdS/CFT~\cite{Maldacena:1997re,Gubser:1998bc,Witten:1998qj,Witten:1998zw,Susskind:1998dq}. There may similarly be  a striking lesson to be learned from the physics of this paper  about the nature of the current observable phase of the  universe, assuming that dark energy's simplest explanation is a positive cosmological constant. In this picture, the ubiquitous black holes in our universe are actually Kerr black holes in a de Sitter universe. Taken as a whole, the peaked specific heat of the system suggests that a microscopic quantum model of our universe should exhibit a fall--off in the degrees of freedom above some highest energy scale. This is in a sense, a precise realization of the idea that de Sitter should, due to the finite cosmological horizon, have some kind of cut--off on the Hilbert space\footnote{This is a feature looked for in both direct constructions of de Sitter vacua such as in ref.~\cite{Dong:2010pm} and in seeking dual descriptions (see {\it e.g.,} refs.~\cite{Spradlin:2001pw,Witten:2001kn}, and references within).}. One way of arguing that was to use the fact that black holes in de Sitter have a maximum size\footnote{In fact, in ref.~\cite{Grumiller:2014oha}, Schottky behaviour was argued in a two dimensional gravitational context, where black holes grow until they approach a cut--off scale.}, and the connection between black hole entropy and microscopic degrees of freedom~\cite{Banks:2000fe}.  The same reasoning might well apply to black holes in the early inflationary universe too.  It would be very interesting to pursue this further to see if it has useful consequences.

\section{Kerr and Reissner--Nordstr\"om de Sitter Black Holes}
\label{sec:Kerr-RN-dS}
\begin{figure}[h]
\centering
\subfigure[]{\centering\includegraphics[width=0.48\textwidth]{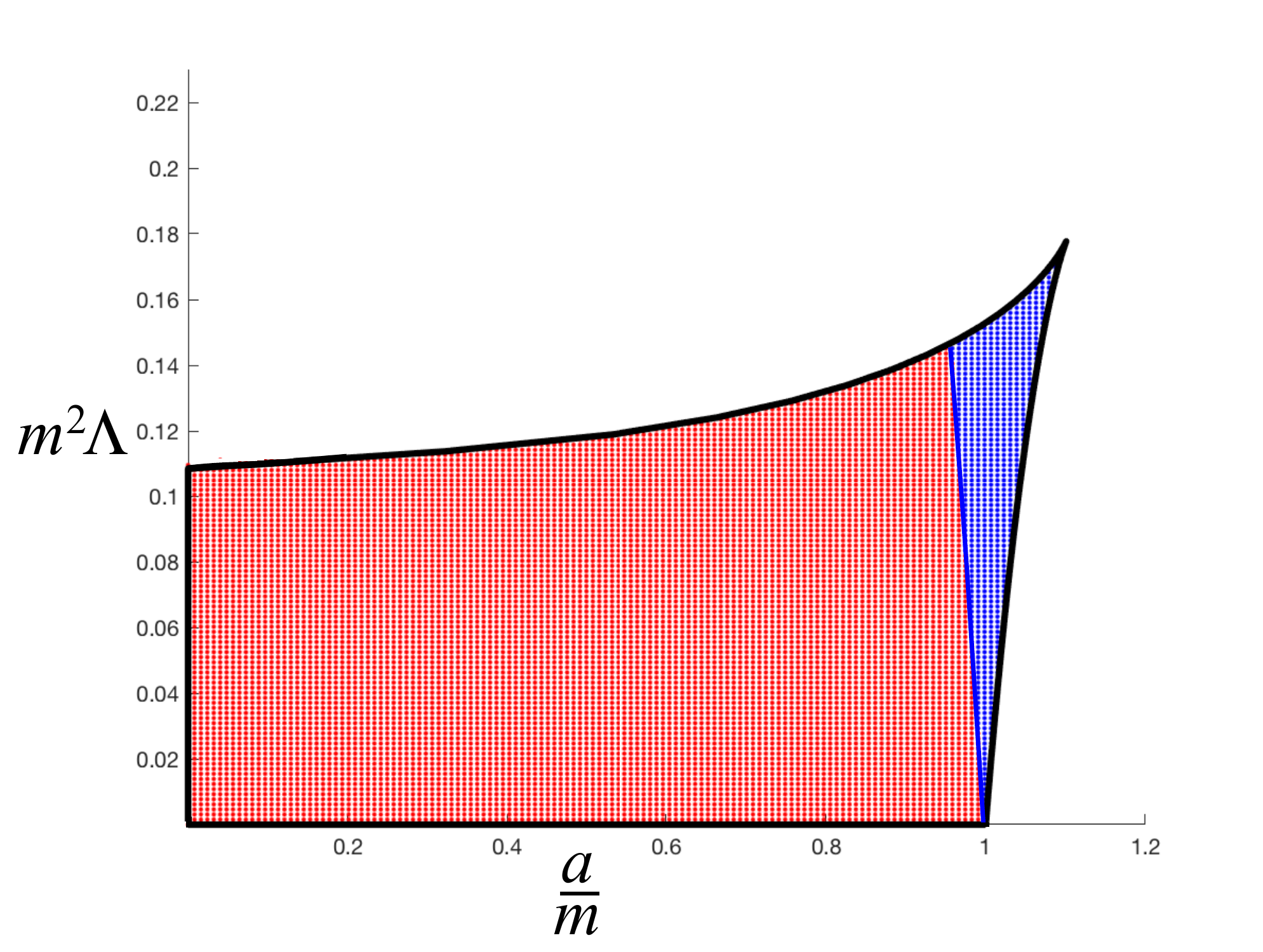}
\label{fig:parameter-space-Kerr-dS}}
\subfigure[]{\centering\includegraphics[width=0.48\textwidth]{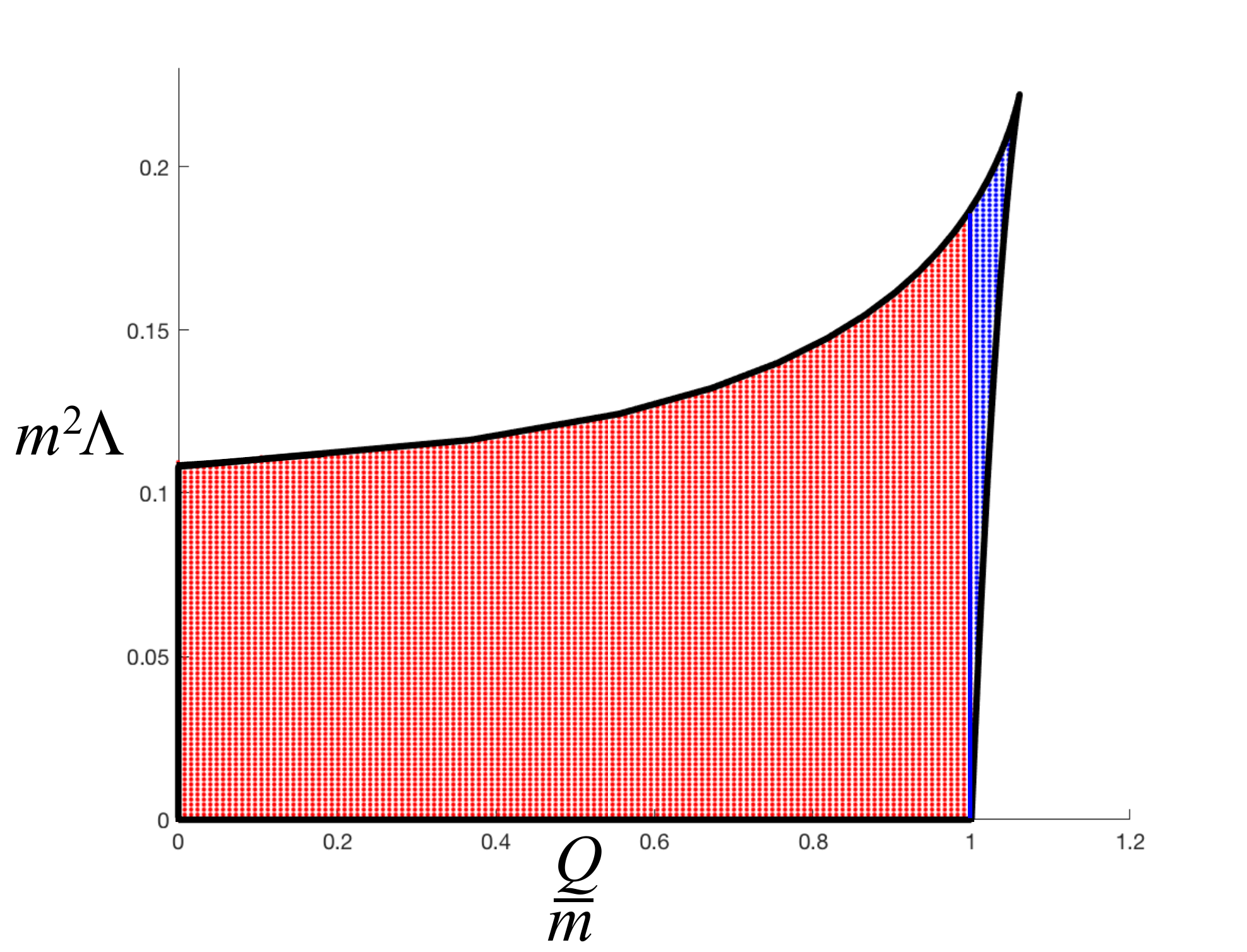}
\label{fig:parameter-space-RN-dS}}
\caption{\label{fig:parameter-space}(a) The parameter space for Kerr--de Sitter  (b) The parameter space for Reissner--Nordstr\"om--de Sitter}
\end{figure}

\begin{figure}[h]
\centering
\subfigure[]{\centering\includegraphics[width=0.48\textwidth]{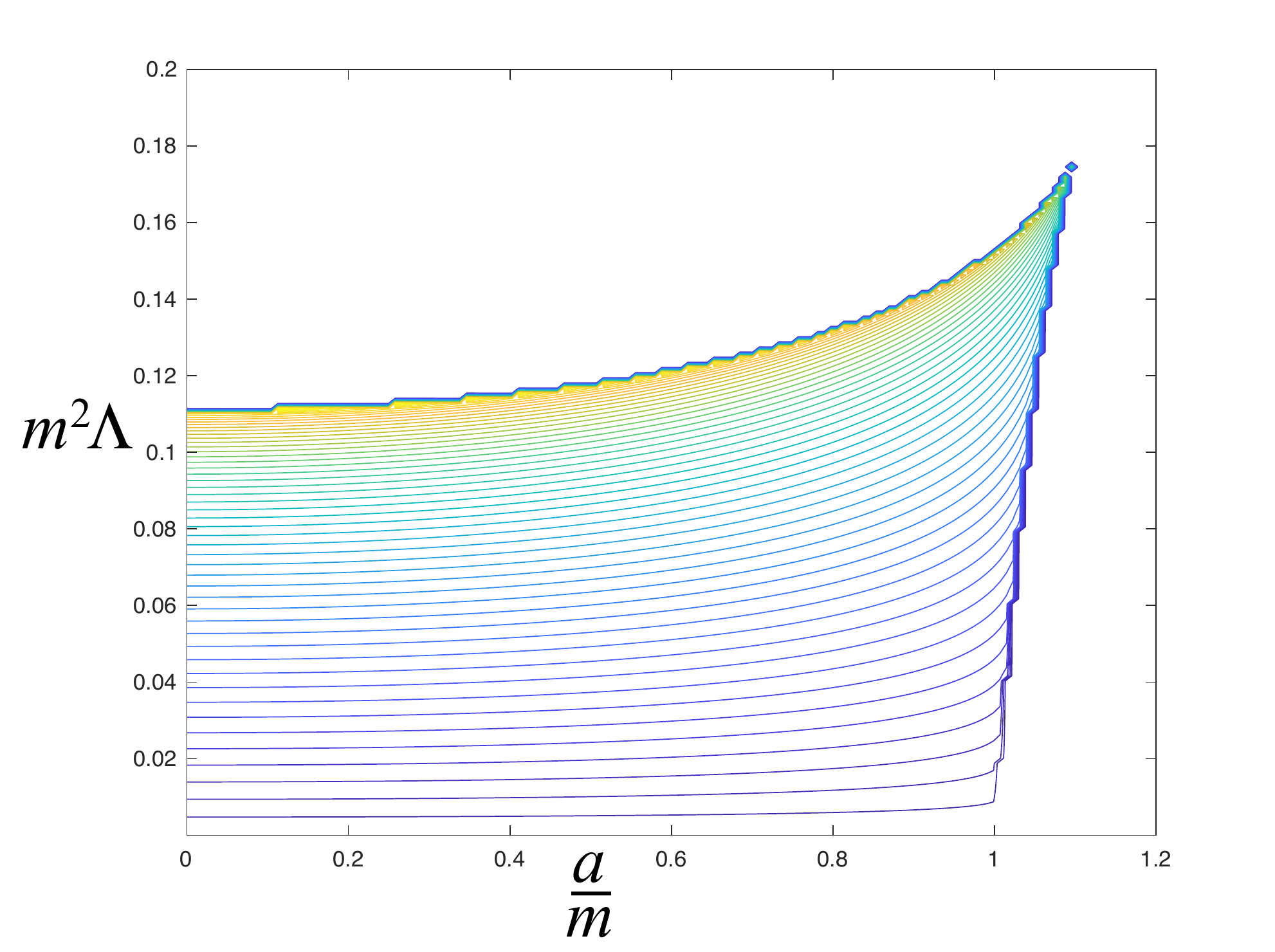}
\label{fig:entropy-bh-Kerr-dS}}
\subfigure[]{\centering\includegraphics[width=0.48\textwidth]{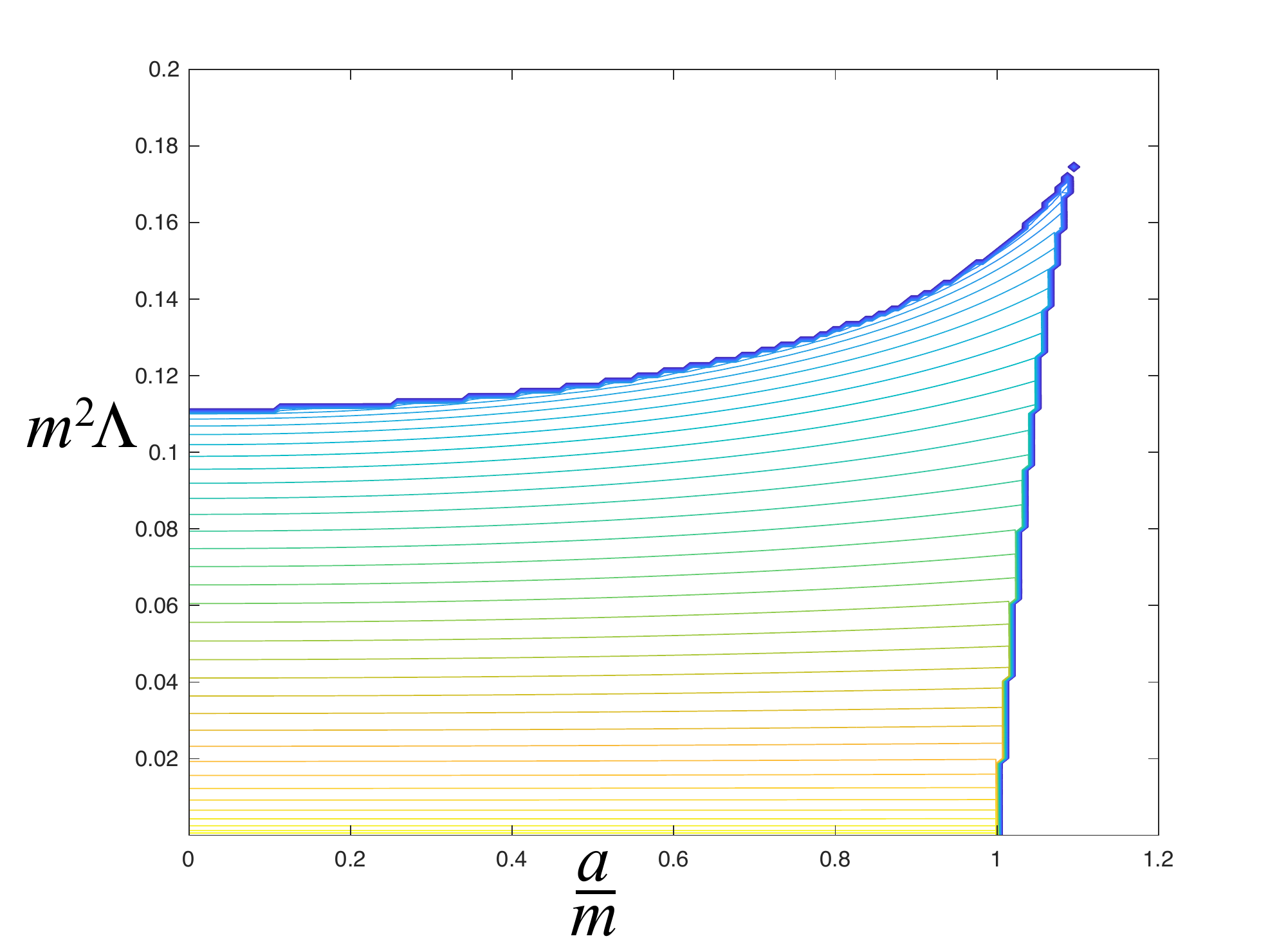}
\label{fig:entropy-c-Kerr-dS}}
\caption{(a) Lines of constant entropy for the black hole in Kerr--de Sitter  (b) Lines of constant entropy for the cosmological horizon in Kerr--de Sitter.}
\end{figure}

\begin{figure}[h]
\centering
\subfigure[]{\centering\includegraphics[width=0.48\textwidth]{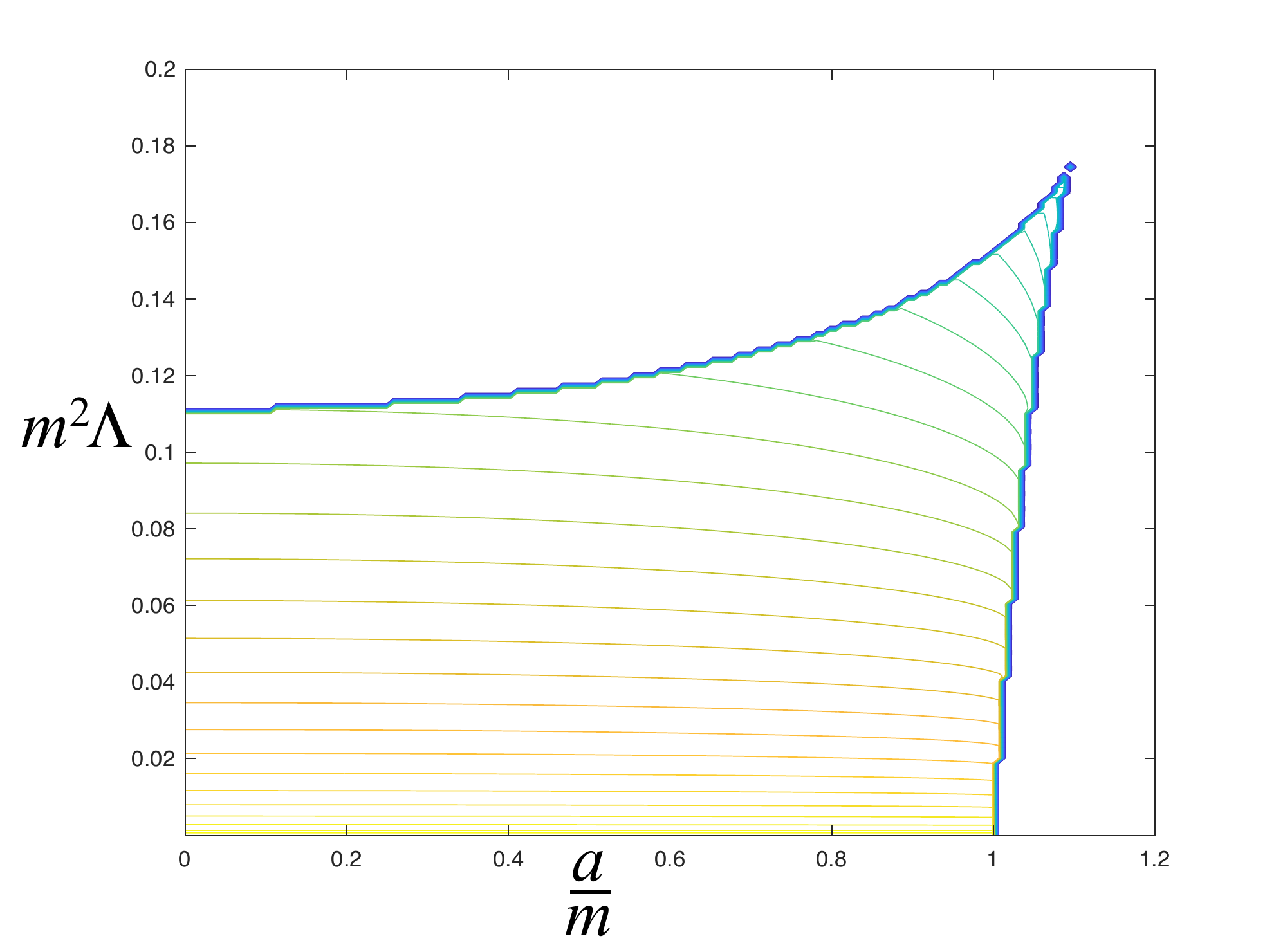}
\label{fig:entropy-bh-c-Kerr-dS}}
\subfigure[]{\centering\includegraphics[width=0.48\textwidth]{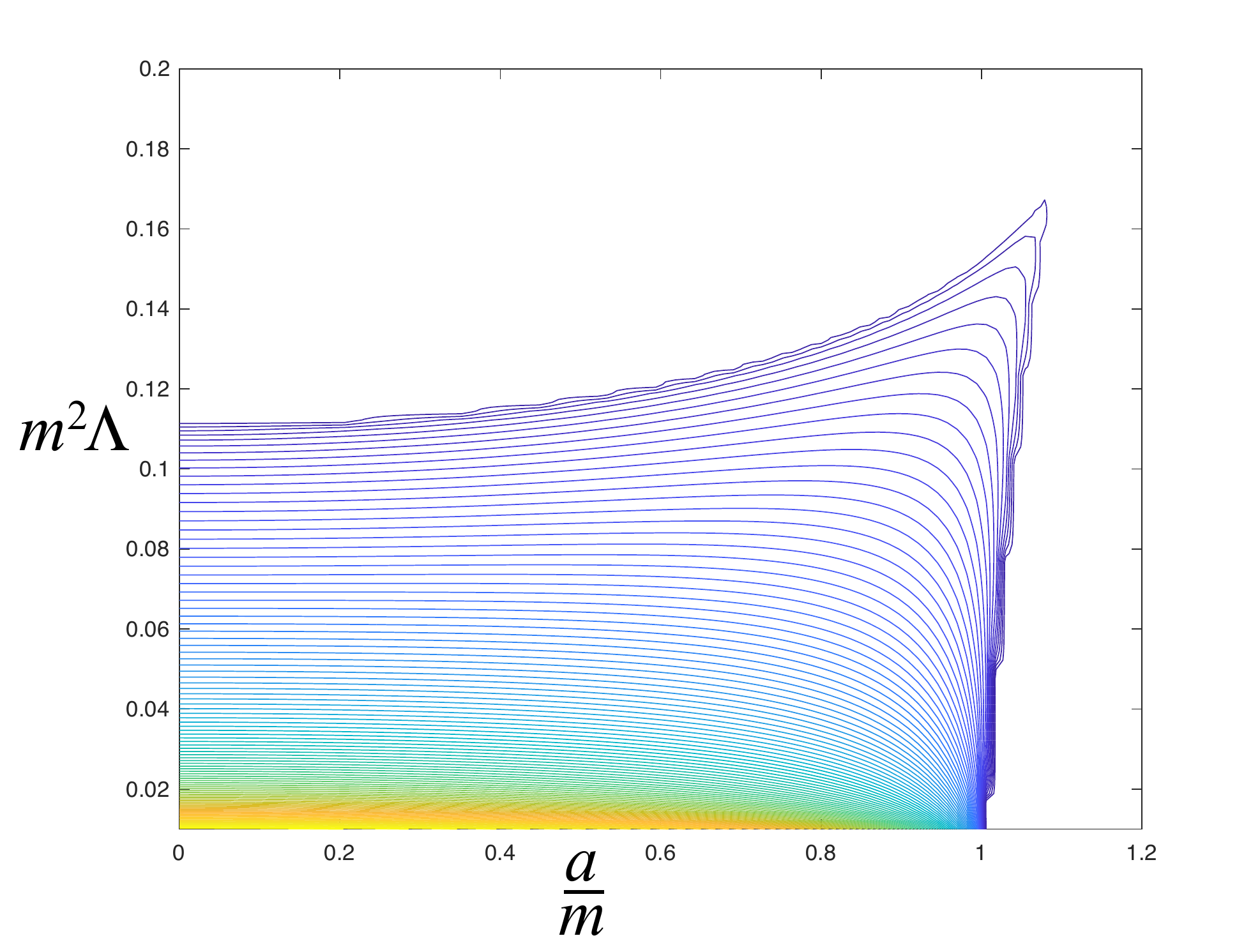}
\label{fig:isotherms-b-Kerr-dS}}
\caption{(a) Lines of constant total entropy for  Kerr--de Sitter  (b) Black hole isotherms in Kerr--de Sitter}
\end{figure}

The Kerr--de Sitter (K--dS) spacetime has metric~\cite{Carter:1968ks,Plebanski:1976gy}:
\begin{eqnarray}
ds^2&=&-\frac{\Delta_r}{\rho^2}\left(dt-\frac{a\sin^2\theta}{\Xi}d\phi\right)^2 +\frac{\rho^2}{\Delta_r}dr^2+\frac{\rho^2}{\Delta_\theta}d\theta^2+\frac{\Delta_\theta\sin^2\theta}{\rho^2}\left(adt-\frac{r^2+a^2}{\Xi}d\phi\right)^2\ ,\nonumber\\
&&{\rm with}\quad\Delta_r\equiv \frac{(r^2+a^2)(\ell^2-r^2)}{\ell^2}-2mr\ ,\quad \Delta_\theta\equiv1+\frac{a^2}{\ell^2}\cos^2\theta\ , \nonumber\\
&& {\rm and}\quad \rho^2\equiv r^2+a^2\cos^2\theta\ , \quad \Xi\equiv1+\frac{a^2}{\ell^2}\ .
\end{eqnarray}
The (positive) cosmological constant is $\Lambda=3/\ell^2$. For the physical range of parameters (positive real mass and spin parameter $a$), there are four roots of $\Delta_r=0$, the three largest of which are positive, and are, in order, $r_c,r_+,r_-$. The fourth is negative, and is $-(r_c+r_++r_-)$. The largest root $r_c$ gives the position of the cosmological horizon while $r_+$ is the position of the black hole horizon. The quantities:
\begin{equation}
\label{eq:physical-kerr}
M=\frac{m}{\Xi^2}\ , \quad  {\rm and}\quad  J=aM
\end{equation}
 are the physical mass and angular momentum, respectively. 
 The temperatures of the black hole and cosmological horizons will be denoted $T_b$ and $T_c$ respectively, and are:
 \begin{equation}
 T_b=\frac{\Delta_r^\prime(r_+)}{4\pi \Xi (r_+^2+a^2)}\ ,\qquad T_c=\frac{\left|\Delta_r^\prime(r_c)\right|}{4\pi \Xi (r_c^2+a^2)}\ ,
 \end{equation}
 and the angular velocities at each horizon are, respectively:
 \begin{equation}
 \Omega_b = \frac{a\Xi}{(r_+^2+a^2)}\ , \qquad \Omega_c = \frac{a\Xi}{(r_c^2+a^2)}\ .
 \end{equation}
 The entropy of an horizon is a quarter of its area, as usual:
  \begin{equation}
S_b = \frac{\pi(r_+^2+a^2)}{\Xi}\ , \quad{\rm and}\quad S_c = \frac{\pi(r_c^2+a^2)}{\Xi}\ .
 \end{equation}
On the other hand, the Reissner--Nordstr\"om--de Sitter (RN--dS) spacetime is:
\begin{eqnarray}
ds^2 &=& -\frac{\Delta_r}{r^2} dt^2 +\frac{r^2}{\Delta_r}dr^2 +r^2(d\theta^2+\sin^2\theta d\phi^2)\ , \nonumber\\
&&{\rm with}\quad\Delta_r\equiv \frac{r^2(\ell^2-r^2)}{\ell^2}-2mr+Q^2\ . 
\end{eqnarray}
Again, there are four zeros of $\Delta_r$, three positive ones 
and a negative one (using the same notation as before). 
In this case,
 \begin{equation}
 T_b=\frac{\Delta_r^\prime(r_+)}{4\pi r_+^2}\ ,\qquad T_c=\frac{\left|\Delta_r^\prime(r_c)\right|}{4\pi r_c^2}\ ,
 \end{equation}
 the entropies of the black hole and of the cosmological horizons are:
$S_b = \pi r_+^2 $ and 
$S_c = \pi r_c^2$ ,
 and the electric potential at each horizon is, respectively:
 \begin{equation}
 \Phi_b = \frac{Q}{r_+}\ , \qquad \Phi_c = \frac{Q}{r_c}\ .
 \end{equation}
For both spacetimes, there are two classes of special cases, arising from degeneracies between positive zeros. The black hole and cosmological horizons can coincide ($r_c=r_+$), or the black hole's inner and outer horizons coincide, making the black hole extremal ($r_+=r_-$). These give two lines in the $(m,a)$ or $(m,Q)$ parameter spaces. The double zero means that the temperatures are zero on these lines. The two cases can overlap in a triple degeneracy point ($r_c=r_+=r_-$). Taken together with the lines where either $a(Q)$ or $m$ vanish, this forms a connected corner of parameter space, which is displayed in figures~\ref{fig:parameter-space-Kerr-dS} and~\ref{fig:parameter-space-RN-dS}, where the axes chosen are $(a/m,m^2\Lambda)$ for K--dS and $(Q/m,m^2\Lambda)$ for RN--dS. (Note that all figures in this paper have $\ell=1$.) The part of parameter space decorated with red dots (on the left in each figure) has $T_b>T_c$ while $T_b<T_c$ for blue (on the right). The line separating the two regions, with $T_b=T_c$ is the so--called ``lukewarm''~\cite{Romans:1991nq} case\footnote{See {\it e.g.} refs.~\cite{Anninos:2010gh,Bhattacharya:2017scw} for reviews of the parametrization of these two spacetimes. See also ref.~\cite{McInerney:2015xwa} for an alternative parametrization of Kerr--Newmann--de Sitter in which the total entropy plays a prominent role.}.
 Crucially, in each case, Kerr--dS or Reissner--Nordstr\"om--dS,  the total entropy of the system  is $S=S_b+S_c$, and plays a central role in the thermodynamics~\cite{Davies:2003me,Urano:2009xn,Saida:2009up,Dolan:2013ft,Bhattacharya:2015mja}.  The specific heat's temperature dependence (with respect to $T_b$, thinking of the black hole as a dynamical physical object placed into de Sitter spacetime) will be our focus, as described in the introduction\footnote{Another reason to study the $T_b$ dependence is that $T_c$ never covers the full range from $0$ to $\infty$, while $T_b$ does (for a broad set of parameters). Being able to study $C(T)$ over a wide range of temperatures is crucial to its utility as a probe of the fundamental underlying degrees of freedom. See section~\ref{sec:schottky-matters}.}. In the quest to understand it, it is interesting to look globally at the parameter space to see how certain physical quantities change. Overall, it is difficult to extract analytical expressions at arbitrary points in parameter space, and temperature derivatives are particularly difficult. 
A numerical strategy (similar to that used in ref.~\cite{Johnson:2019vqf} to extract $C(T)$) was adopted. The  dots in  figure~\ref{fig:parameter-space}  are sample points (a grid of 150$\times$150) at which the physical quantities like $(S_b,S_c,J,M,T_a,T_b)$ were computed numerically and stored. Then the data set could be mined for various quantities such as the specific heat  along chosen curves of interest. From those data sets, for example, we can  examine sample curves of constant entropy for the black hole horizon and for the cosmological horizon, as shown in figure~\ref{fig:entropy-bh-Kerr-dS} and~\ref{fig:entropy-c-Kerr-dS}. Curves of constant total entropy are shown in figure~\ref{fig:entropy-bh-c-Kerr-dS}, while the~$T_b$ isotherms are shown in figure~\ref{fig:isotherms-b-Kerr-dS}. %
Care must be taken in reading these figures near the edges, since the accumulation of curves becomes more prone to numerical error. A lot can be learned from these figures (similar ones were made for RN--dS). Starting first with the isotherms, recall that $T_b=0$ on the extremal  line (the rightmost edge) and on the upper edge, where the black hole is as large as it can be (for a given $a/m$). The black hole's entropy is largest there, while that of the cosmological horizon is smallest (again, at a given $a/m$). 

\begin{figure}[h]
\centering
\subfigure[]{\centering\includegraphics[width=0.48\textwidth]{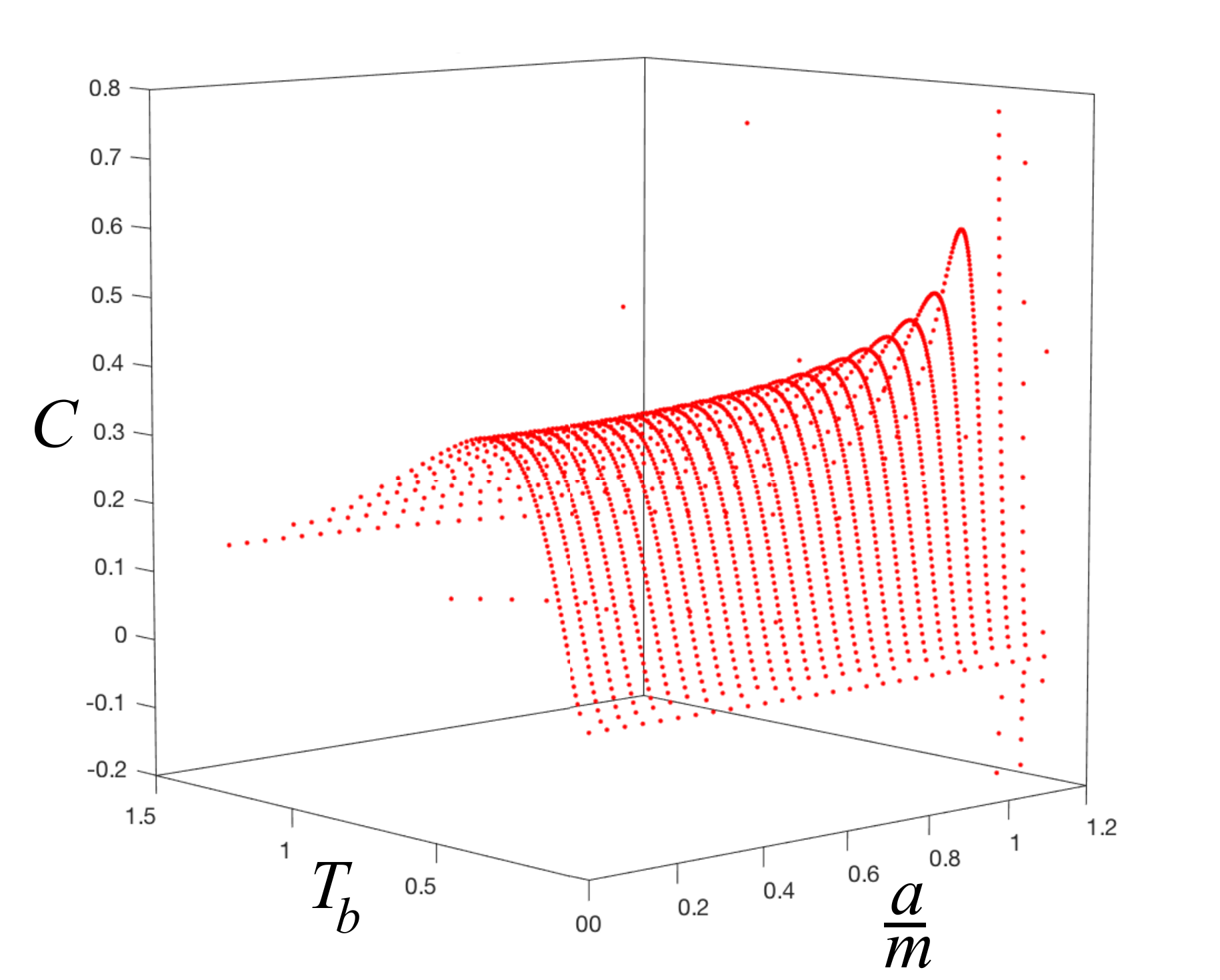}
\label{fig:specific-heat-aovrm-Kerr-dS}}
\subfigure[]{\centering\includegraphics[width=0.465\textwidth]{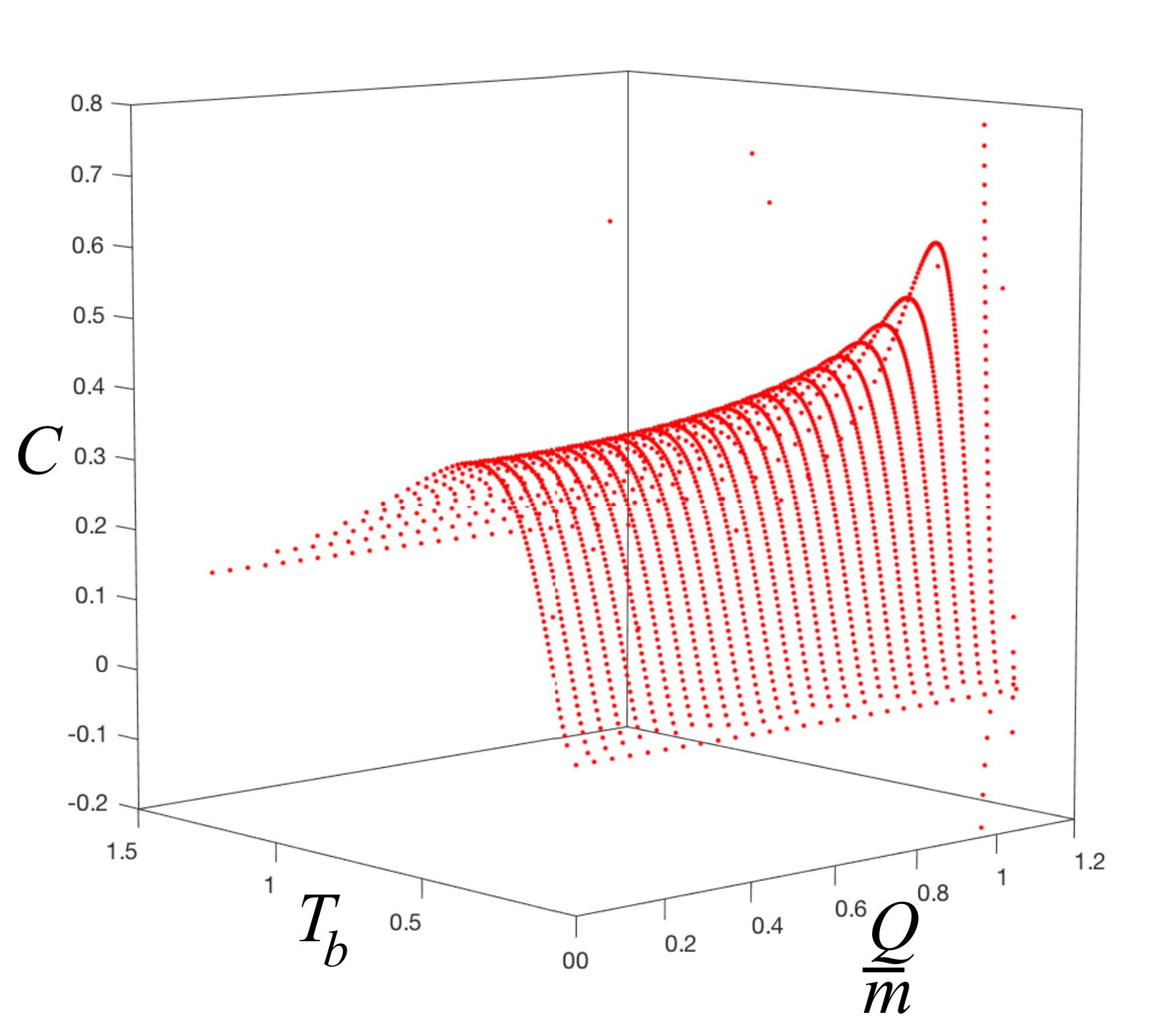}
\label{fig:specific-heat-Qovrm-RN-dS}}
\caption{(a) Slices of the specific heat $C(T_b)$ for  Kerr--de Sitter. For $a/m<1$ $C(T_b)$ displays Schottky--like behaviour.  (b) Slices of the specific heat $C(T_b)$ for  Reissner--Nordstr\"om--de Sitter. For $Q/m<1$ $C(T_b)$ displays Schottky--like behaviour.  }
\end{figure}

\noindent Turning to the specific heat:
\begin{equation}
\label{eq:specific-heat}
C=T_b\frac{\partial S}{\partial T_b}\ ,
\end{equation}
there are several options as to what slices in parameter space to compute it along. The most straightforward is along lines of constant $a/m$ (or $Q/m$), vertical slices in the parameter space displayed so far. Lines for $a/m<1$  ($Q/m<1$) will cover the whole range of temperatures $0<T_b<\infty$, resulting in a single branch for $C(T_b)$, while lines for $a/m\ge1$ will start at $T_b=0$, rise to a maximum temperature, and then return to $T_b=0$. This will result in two branches for~$C(T_b)$. Figure~\ref{fig:specific-heat-aovrm-Kerr-dS} shows several curves for $C(T_b)$ at different $a/M$, with the equivalent figure for RN--dS in figure~\ref{fig:specific-heat-Qovrm-RN-dS}. For $a/m<1$ ($Q/m<1$) there is a family of Schottky--like peaks, while for $a/m\ge1$ ($Q/m\ge1$) (there are two cases visible) there is only a part of the function present due to the expected truncation at a maximum temperature. There is a second branch of $C(T_b)$ visible, which is negative. (A closer look at this kind of behaviour can be seen by considering constant $J$, next.)

\begin{figure}[h]
\centering
\subfigure[]{\centering\includegraphics[width=0.49\textwidth]{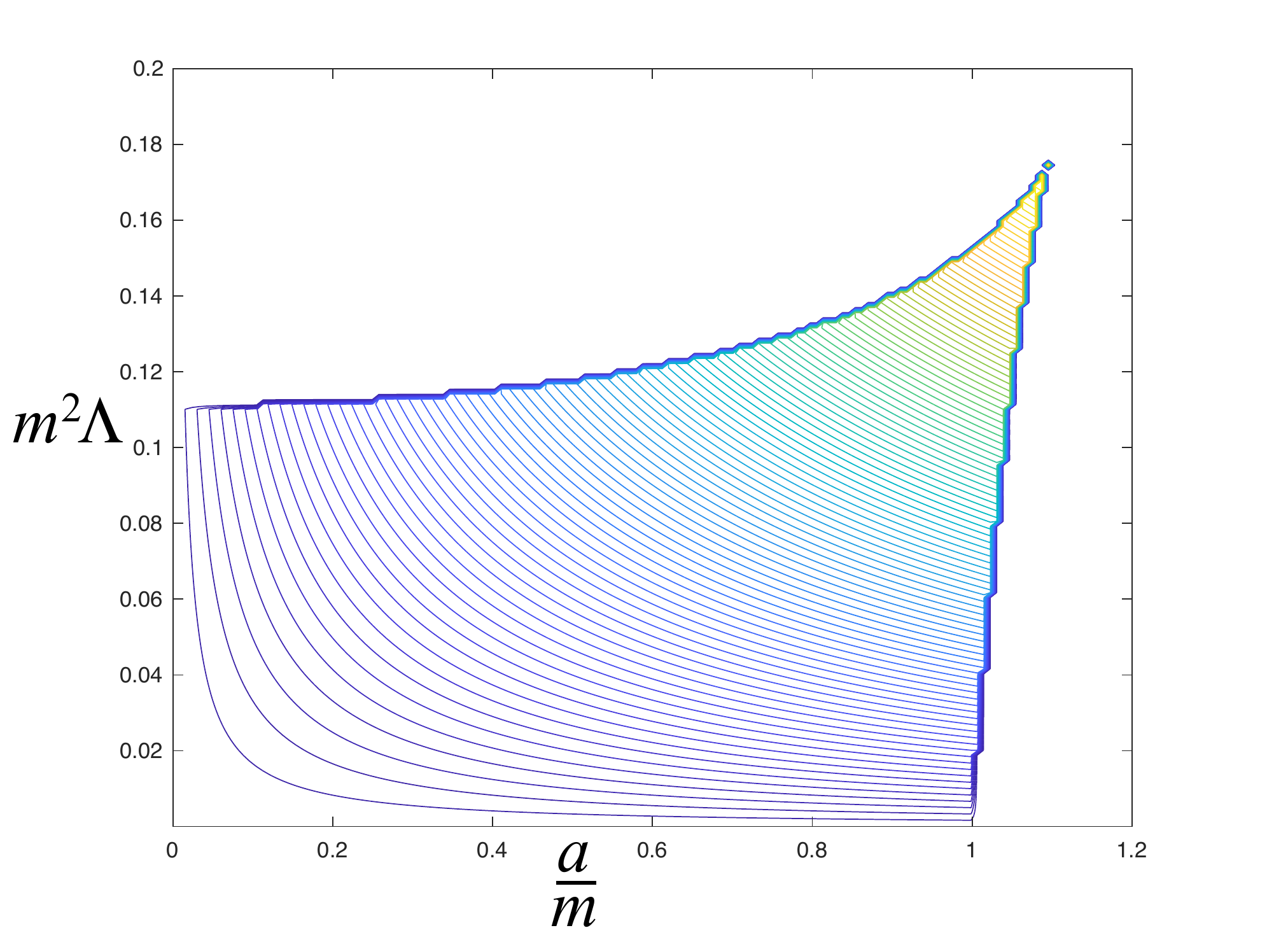}
\label{fig:angmom--Kerr-dS}}
\subfigure[]{\centering\includegraphics[width=0.45\textwidth]{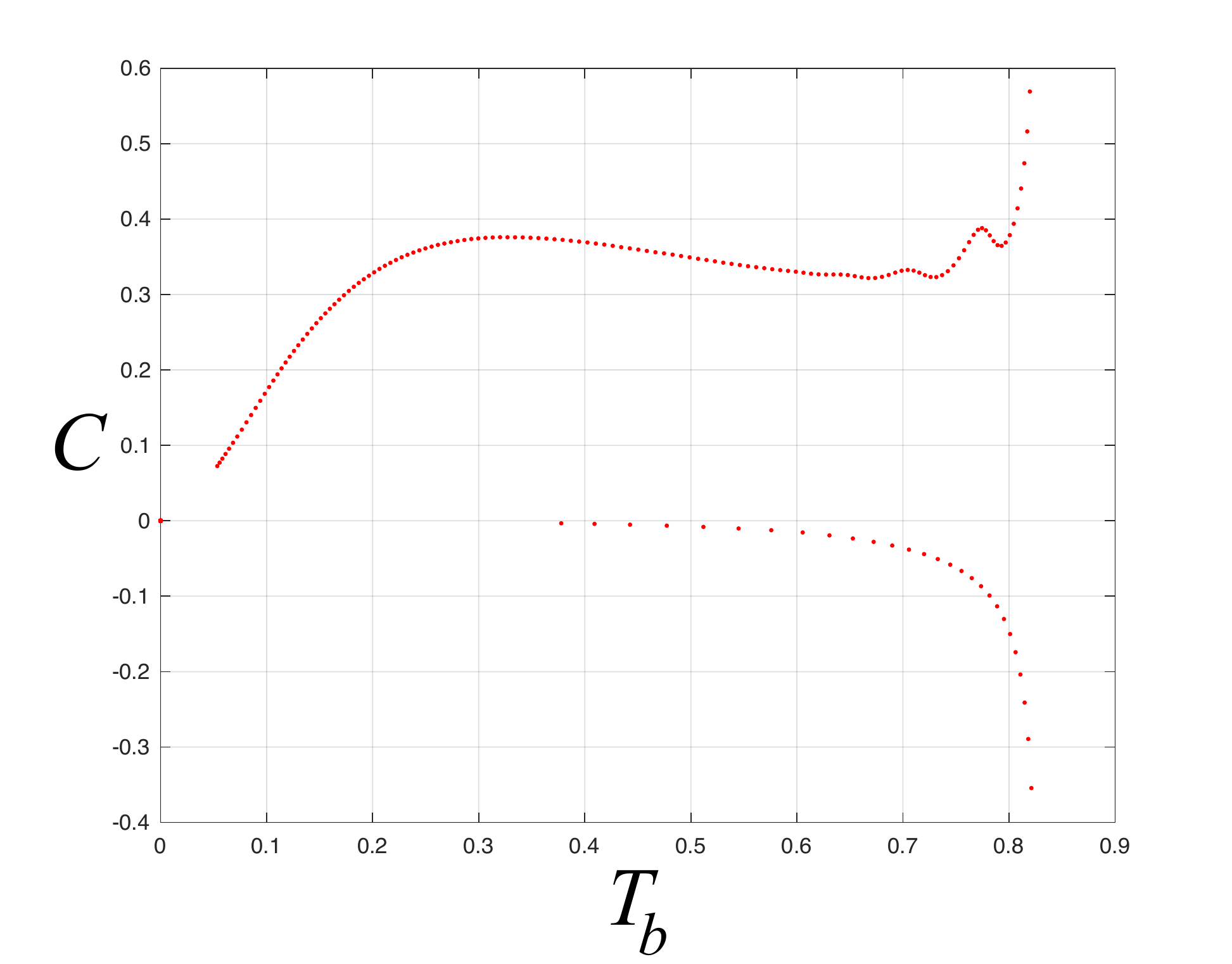}
\label{fig:specific-heat-constant-J--Kerr-dS}}
\caption{(a)  Lines of constant angular momentum $J$ in Kerr--de Sitter.  (b) Partial Schottky--like peak for constant $J\simeq1.1\times10^{-3}$ in Kerr--de Sitter, along with the truncation at a maximum temperature, $T^J_{\rm max}\simeq0.82$.  }
\end{figure}

Other options include computing $C(T_b)$ at fixed $J$ (or fixed $Q$ for RN--dS). In such cases, again the full range of temperature is not accessible. Lines of constant  $J$  (or $Q$) start out at $T_b=0$ (the extremal line) and return to $T_b=0$ again at large $m$. See figure~\ref{fig:angmom--Kerr-dS}.  A maximum temperature is reached in between. Again, this results in two branches for $C(T_b)$ for the range of temperatures that can be reached. Part of the Schottky peak can  appear, depending upon how high a temperature $T^J_{\rm max}$ can be achieved on that curve\footnote{This is reminiscent of the Schottky--like story uncovered for Kerr--AdS and STU--AdS (for constant volume) in ref.~\cite{Johnson:2019vqf}. There, depending upon the value of $J$ or $Q$, the peaks could sometimes be hidden by other structures, the region of the critical point those systems have.}. See figure~\ref{fig:specific-heat-constant-J--Kerr-dS} for  an example. (Some jaggedness of the figure is due to loss of numerical accuracy, but the core features---the peak and the truncation at a maximum temperature---should be clear.)

\section{Thermodynamics of Schottky--like Peaks}
\label{sec:schottky-matters}

\begin{figure}[h]
\centering
\subfigure[]{\centering\includegraphics[width=0.32\textwidth]{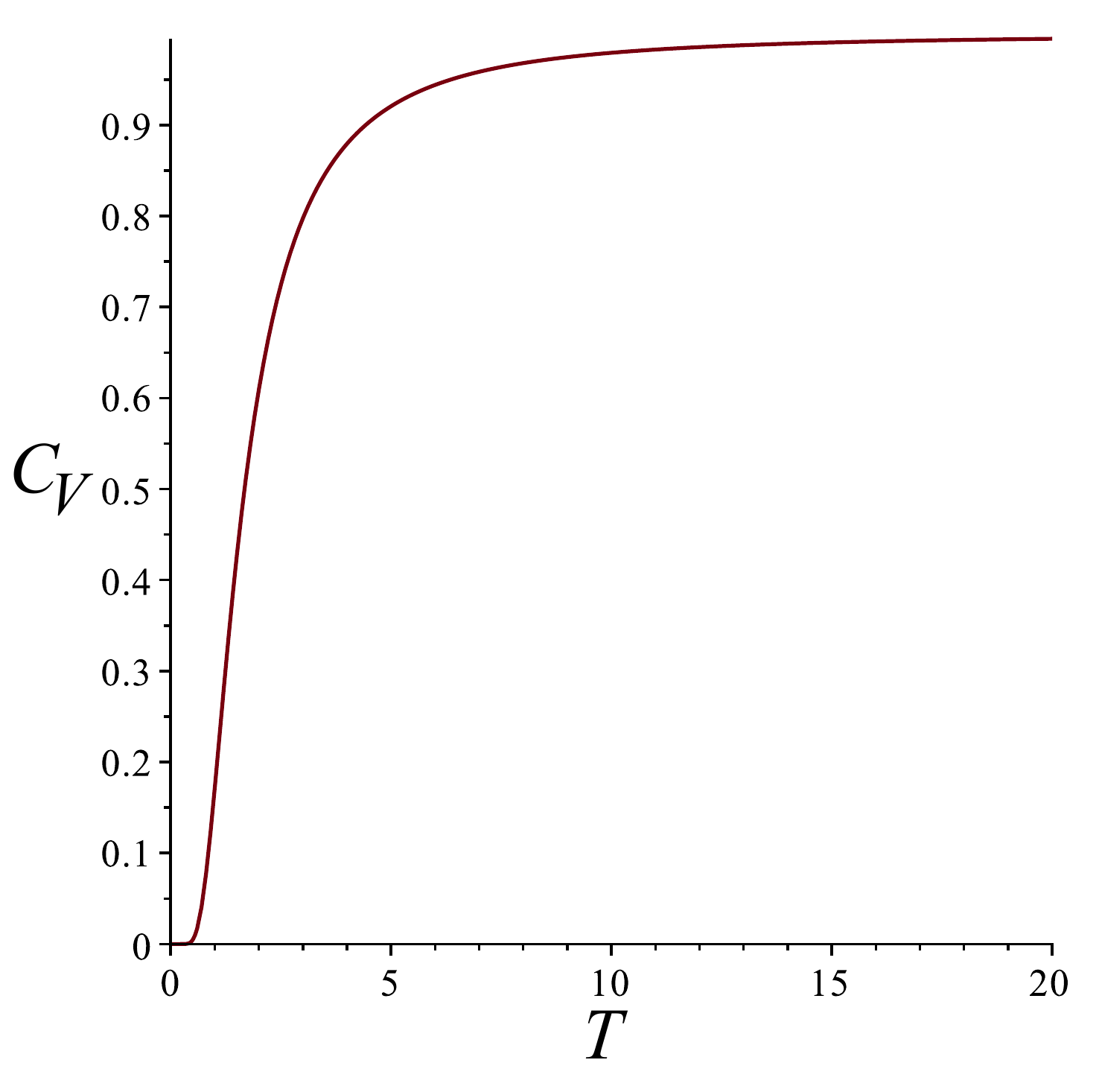}
\label{fig:QHO-Cv}}
\subfigure[]{\centering\includegraphics[width=0.32\textwidth]{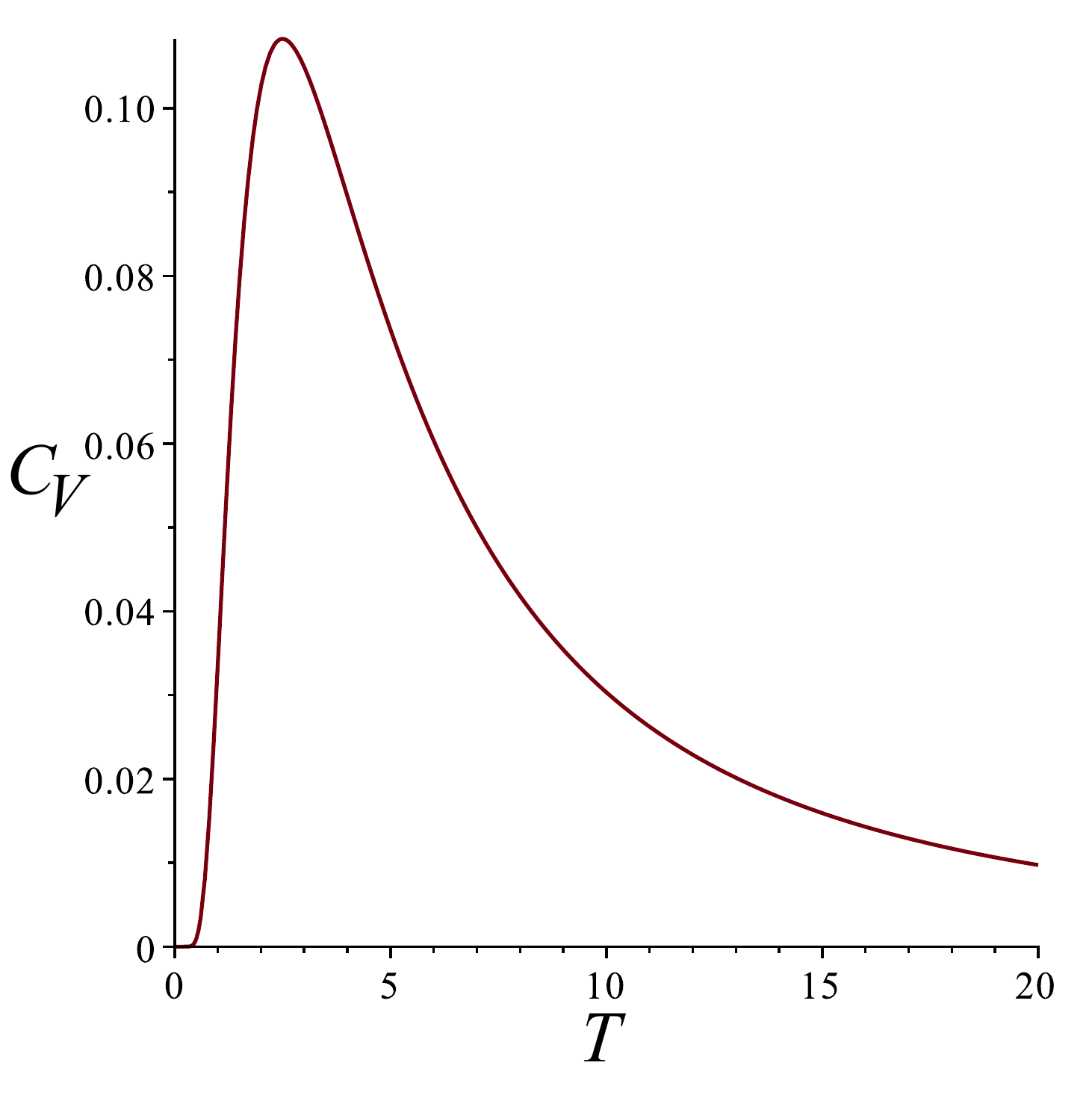}
\label{fig:QHO-truncated-Cv}}
\subfigure[]{\centering\includegraphics[width=0.32\textwidth]{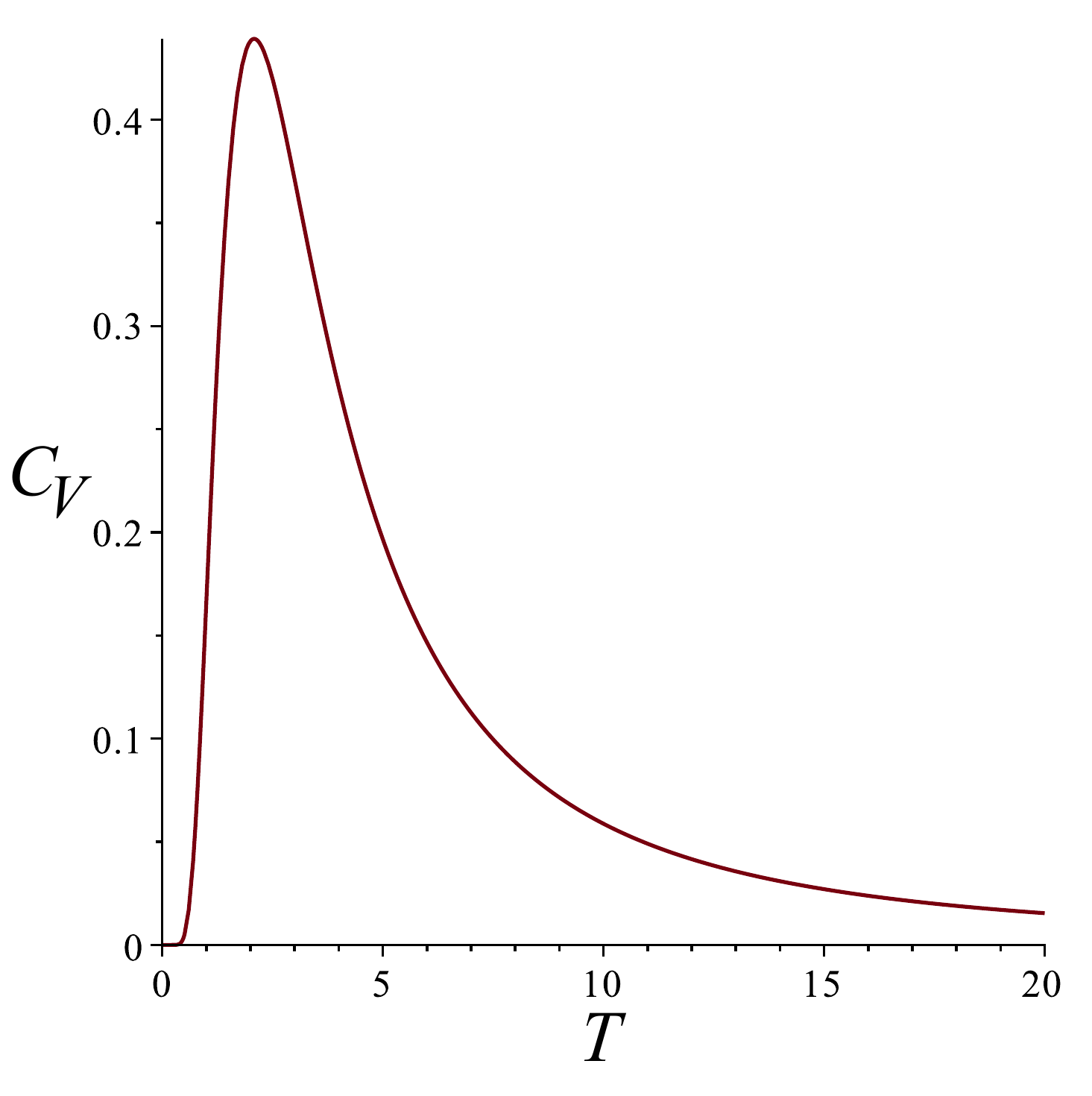}
\label{fig:Schottky-Cv}}
\caption{Three examples of the temperature dependence of the specific heat $C_V(T)$ (a) The  quantum harmonic oscillator, with $\Delta=5$. (b) The quantum harmonic oscillator truncated to $n=5$. Here, $\Delta=1$. (c) A two--level Schottky model, with $\Delta=5$.}
\end{figure}

 As reviewed in ref.~\cite{Johnson:2019olt}, while the classic Schottky peak  arises from simple two (or few) level systems (such as spins or defects), a Schottky--like peak can appear as a result of truncating a system such that the degrees of freedom cannot access higher energies. This is easily modelled using the quantum harmonic oscillator, which has  energies $E_n=(n+\frac12)\hbar\omega$. Its specific heat is (the classic 1907 Einstein model~\cite{Einstein1907}): 
\begin{equation}
C_V^{\rm QHO}=k_B\left(\frac{\Delta}{k_BT}\right)^2\frac{{\rm e}^{\beta\Delta}}{(1-{\rm e}^{\beta\Delta})^2}\ , \qquad {\rm where}\quad \beta=\frac{1}{k_B T}\ ,\quad \Delta=\hbar\omega\ .
\end{equation}
It is displayed in figure~\ref{fig:QHO-Cv}. 
This specific heat has a representation as an infinite series:
\begin{equation}
C_V^{\rm QHO}=k_B\left(\frac{\Delta}{k_BT}\right)^2\sum_{n=1}^\infty n {\rm e}^{-n\beta\Delta}\ ,
\end{equation}
and a truncation to a finite $n$  can be done, effectively turning off the contributions above a chosen excitation level. This results in a peak such as the one shown in figure~\ref{fig:QHO-truncated-Cv}. This is qualitatively very similar to the classic Schottky specific heat for a two level system:
\begin{equation}
C_V^{\rm Schottky}= k_B\left(\frac{\Delta}{k_BT}\right)^2\frac{{\rm e}^{\beta\Delta}}{(1+{\rm e}^{\beta\Delta})^2} =  k_B\left(\frac{\Delta}{k_BT}\right)^2\sum_{n=1}^\infty  (-1)^{n-1} n {\rm e}^{-n\beta\Delta}\ ,
\end{equation}
where here $\Delta$ is the energy gap between the levels. This is depicted in figure~\ref{fig:Schottky-Cv}.  The point of repeating this here is to emphasize that the most important/robust feature signalled by the Schottky peaks is that there is some kind of maximum available energy for the underlying degrees of freedom. It is this notion that there is a maximum energy, more than anything else, that will be taken forward into the next sections.

\section{Thermodynamics with Two Baths}
\label{sec:two-baths}
Part of the challenge of understanding physics in de Sitter spacetimes is the tension between writing down equilibrium thermodynamic quantities and the fact that the system is clearly not in equilibrium. Progress can be made, however.

First, it is worthwhile summarizing the key features. There are two natural temperatures associated to these spacetimes, one associated to the black hole horizon and the other  to the cosmological horizon. For most of the parameter space, the black hole has the higher temperature, although there is an interesting corner where it is the cosmological spacetime that has the higher temperature. In either of those regimes, the system is unstable, with the higher temperature sector depositing energy into the lower temperature sector. 

\begin{figure}[h]
\centering
\subfigure[]{\centering\includegraphics[width=0.44\textwidth]{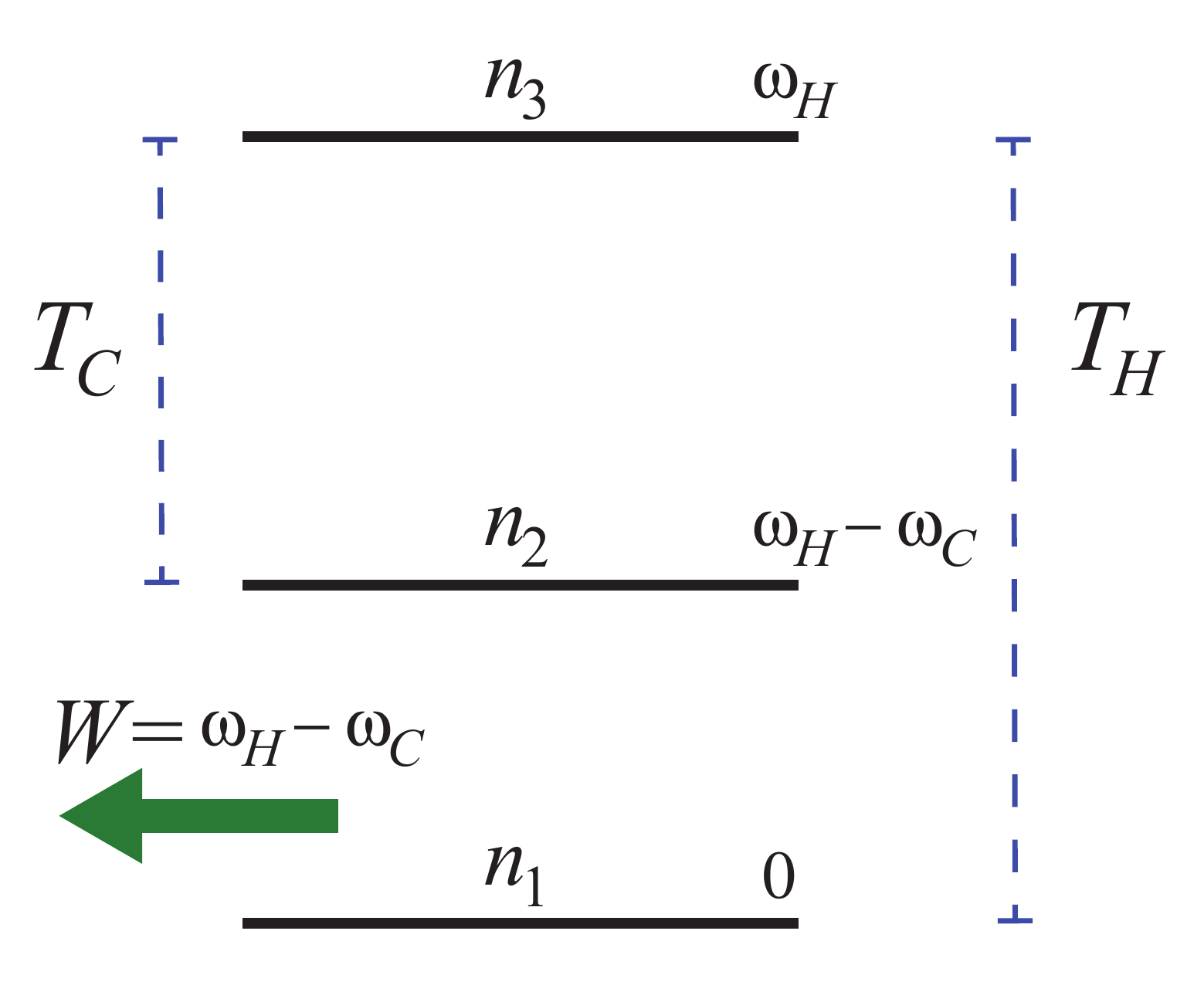}
\label{fig:maser-i}}\hskip 1.0cm 
\subfigure[]{\centering\includegraphics[width=0.44\textwidth]{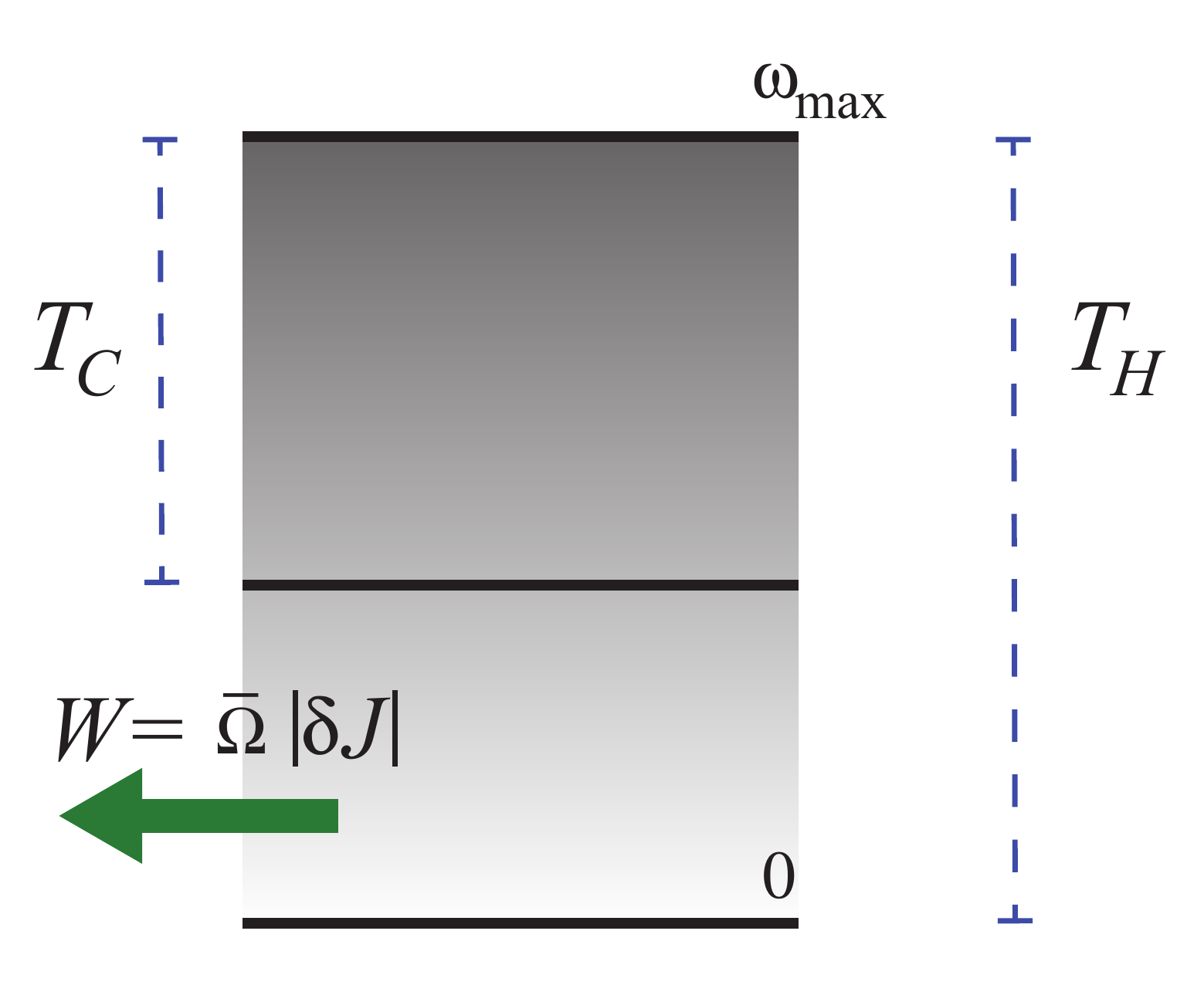}
\label{fig:maser-ii}}
\caption{(a)  The classic three level maser as a quantum continuous heat engine.  (b) de Sitter black holes as a continuous heat engine. Key features shared with the maser are a highest energy, $\omega_{\rm max}$, and a work output channel, which involves loss of angular momentum (for Kerr--dS). Here, ${\bar\Omega}\equiv \Omega_b+\Omega_c-2\Omega_\infty$.  }
\end{figure}

There is another  natural kind of  instability that is present when there are two temperatures, and it is familiar from some of the earliest models of quantum thermodynamics of systems with a finite energy range. The simplest model uses three energy levels, but  {\it the number of levels  is not as important as the presence of the  finite energy cap}.  Imagine connecting the system to a heat bath at temperature $T_H$, exciting states from the ground state to the highest energy $\omega_H$. See figure~\ref{fig:maser-i}. 
States can decay down to some intermediate level $\omega_H-\omega_C$ by giving up some energy to a heat bath at some temperature $T_C<T_H$. There can also be  some transition from the intermediate level back down to the ground state.
Boltzmann distributions determine the average relative filling of energy levels:
\begin{equation}
\label{eq:boltzmann}
\frac{n_3}{n_1}=\exp\left(-\frac{\omega_H}{T_H}\right)\ , \qquad \frac{n_3}{n_2}=\exp\left(-\frac{\omega_C}{T_C}\right)\ .
\end{equation}
Notice that both of these ratios are less than unity,  as per standard equilibrium thermodynamics. The ratio $n_2/n_1$ is what will be of interest. The two cases are:

\begin{itemize}

\item $n_2/n_1<1.$ This is the  ``normal'' state of affairs, and  can simply be described  in terms of a third heat bath in a Boltzmann fashion, with some temperature ${\widetilde T}<T_H$.  On average, there's energy flow  in each of these three channels, and it is appropriately described as heat flow, the sign of which is determined by the ordering of the temperatures: $Q_H=Q_C+{\tilde Q}$. Alternatively, the $T_C$ heat bath can  be expanded to take into account all the heat flow out:
\begin{equation}
\label{eq:just-heat}
Q_H=Q_C\ .
\end{equation}

\item  $n_2/n_1>1$. This interesting situation is now not equilibrium heat flow, but something else. It is work, $W$. (Formally, this  can be described as heat flow at infinite ${\widetilde T}$.) This is population inversion, and can be thought of as an instability of the thermodynamic system not normally seen in equilibrium thermodynamics, but which can be brought about by the interplay of multiple heat baths, as is the case here. Now, the situation is better written as:
\begin{equation}
\label{eq:heat-work}
Q_H=Q_C+W \ .
\end{equation}

\end{itemize}

\noindent All of this amounts to a description of what is called a continuous heat engine, which can last indefinitely if connected to an infinite heat source and heat sink, or for a finite lifetime (long compared to the transition rates between levels) until the inversion goes away.  This is  the basis for a laser or maser, and was first given a quantum thermodynamic interpretation by Scovil and Schulz--Du Bois~\cite{PhysRevLett.2.262}. In fact, the condition for population inversion and maser emission is that the efficiency of the engine be less than or equal to the Carnot efficiency, as will be  discussed below. This three--state model is the original 
{\it quantum} heat engine.

\subsection{de Sitter Black Holes as Continuous Heat Engines}
\label{sec:heat-engine}

The point here is that all the key features needed for the maser action above are present  for de Sitter black holes. There are two temperatures, and as has now been established with the Schottky--like peaks, a notion of a maximum energy per degree of freedom. Discrete finite levels are not needed in order to have the key phenomenon, which is an effective population inversion resulting in an output of work. In the case of Schwarzschild--dS there is no work. The first law of (ordinary) black hole thermodynamics comes in two components~\cite{Dolan:2013ft}:
\begin{equation}
\label{eq:just-heat-i}
dM=T_bdS_b \ , \quad {\rm and}\quad -dM=T_cdS_c
\end{equation}
for the black hole and cosmological horizons, respectively. In other words, the system can increase its mass--energy by heat inflow $T_bdS_b$, an increase of the size of the black hole horizon.  Size increase of the cosmological horizon results in a reduction of mass--energy. This is heat outflow. For no changes in mass, the difference can be written as 
\begin{equation}
\label{eq:just-heat-ii}
T_bdS_b=T_cdS_c \ ,
\end{equation}
which should be compared to equation~(\ref{eq:just-heat}), where $T_b=T_H$ and $T_c=T_C$. (This is if staying to the left of the lukewarm region of parameter space---the temperatures are exchanged otherwise). The total entropy is $S=S_b+S_c$, and in this picture it makes sense that its dependence on the black hole temperature~$T_b$ was natural to examine. The natural ``effective'' energy in this picture is not the black hole energy but that defined by $d{\widehat E}=T_bdS$. It is:
\begin{equation}
{\widehat E}=(1-T_b/T_c)M\ ,
\end{equation}
and the specific heat $C=(1-T_b/T_c)C_b$ where $C_b=T_b\partial S_b/\partial T_b$, the specific heat of the black hole. The latter is negative, and so is the temperature dependent factor\footnote{In fact, this makes ${\widehat E}$ negative, so a better choice might have been to either flip the sign on the specific heat, or use an extra sign in ${\widehat E}$'s definition.}, giving the positive Schottky--like peaks see in previous sections (also recently reported for Schwarzschild--dS in ref.~\cite{Dinsmore:2019elr}).

Things get much richer when the system has either charge or angular momentum. Then there are $\Phi_i dQ$ and $\Omega_i dJ$ ($i=b,c$) terms in the individual first laws for each horizon. The sum gives~\cite{Dolan:2013ft}:
\begin{equation}
\label{eq:energy-heat-work-i}
2dM=T_bdS_b-T_cdS_c +(\Omega_b+\Omega_c-2\Omega_\infty)d J\ ,
\end{equation}
 (where $\Omega_\infty=a/\ell^2$), with an obvious analogue for the charged case. The continuous heat engine mode of operation will be to have  heat and work flows while leaving the energy, $M$, fixed, hence:
\begin{equation}
\label{eq:heat-work-ii}
T_bdS_b=T_cdS_c -(\Omega_b+\Omega_c-2\Omega_\infty)d J\ ,
\end{equation}
This  should be compared to equation~(\ref{eq:heat-work}), where $W=-(\Omega_b+\Omega_c-2\Omega_\infty)dJ$ is to be interpreted as work done.  Situations for which  $dJ=0$  are most analogous to the Schwarzschild--dS case: The black hole, if it starts out with its temperature higher than the cosmological horizon,  stays at fixed mass, with heat $Q_H=T_bdS_b$ flowing in from the hot bath  and an equal amount of heat $Q_C=T_cdS_c$ flowing out to the cold bath. The new work channel is the possibility of doing $dJ<0$ processes. This is the precise analogue of the population inversion work channel in the three level system.

\begin{figure}[h]
\centering
\subfigure[]{\centering\includegraphics[width=0.48\textwidth]{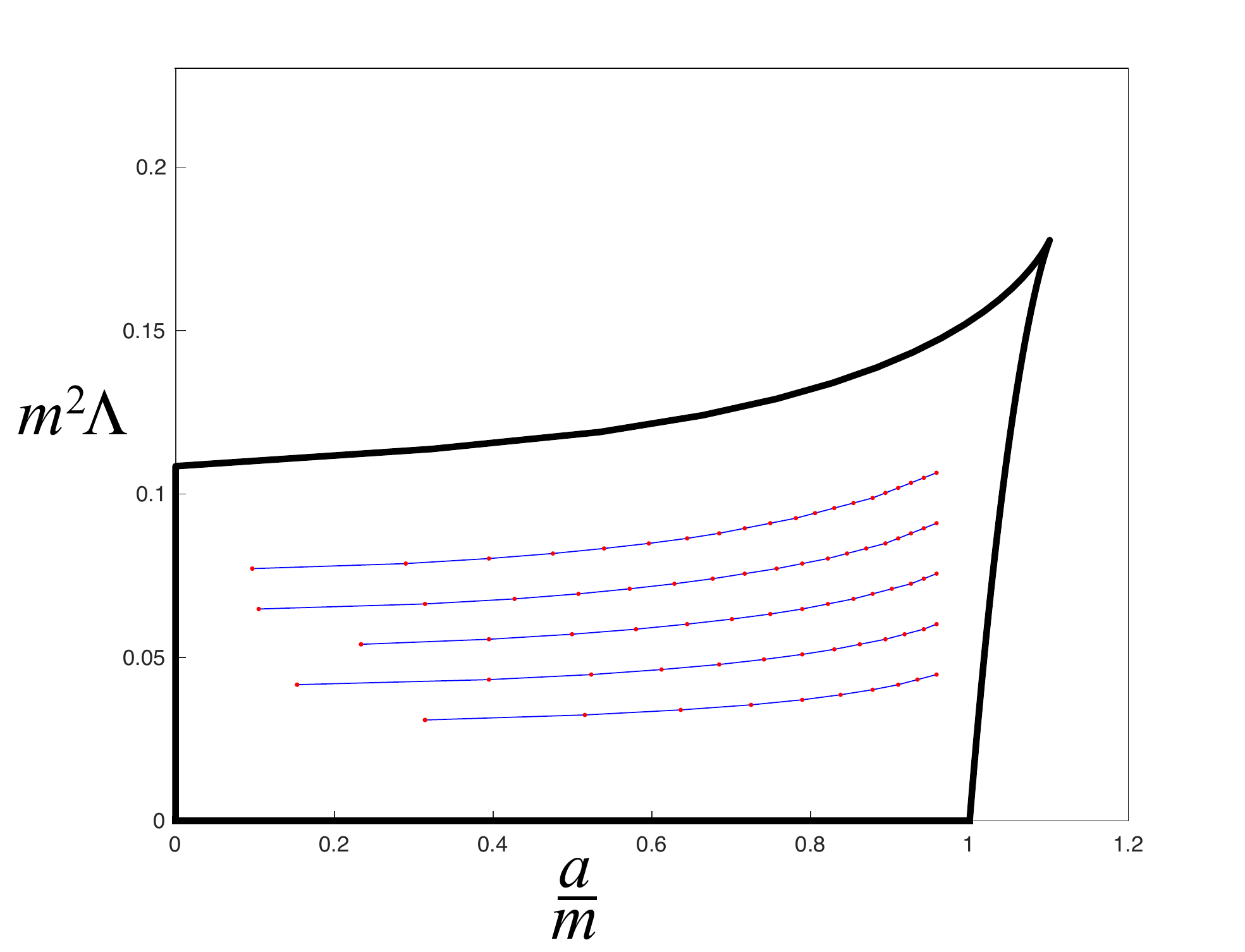}
\label{fig:trajectories-penrose-Kerr-dS}}
\subfigure[]{\centering\includegraphics[width=0.48\textwidth]{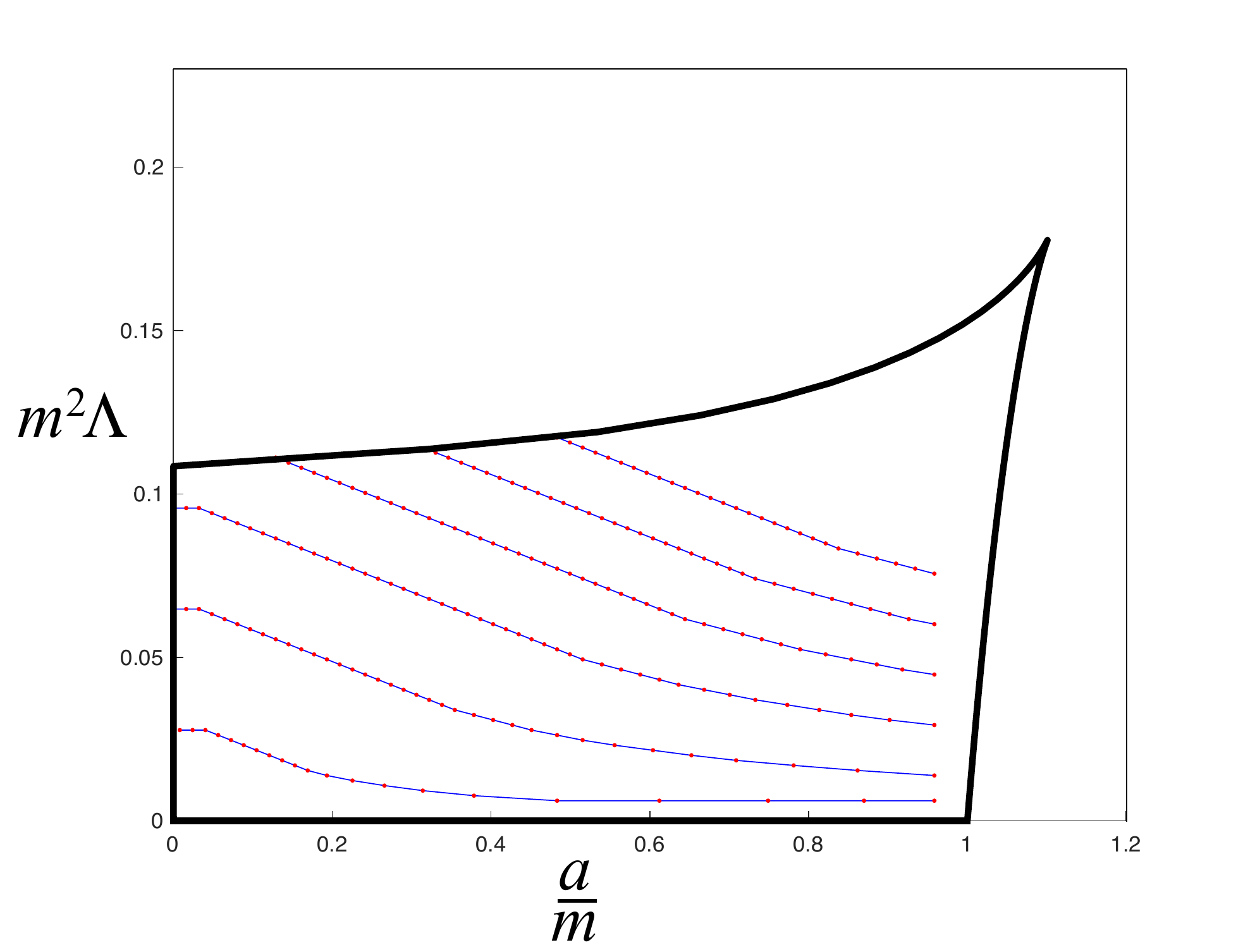}
\label{fig:trajectories-non-penrose-Kerr-dS}}
\caption{Sample numerically determined trajectories  (moving from right to  left) in  Kerr--de Sitter parameter space: (a) Penrose trajectories, where $\delta J<0$ and $\delta M<0$, and (b) Non--Penrose trajectories  where $\delta J<0$ but $\delta M>0$.}
\end{figure}

\begin{figure}[h]
\centering
\subfigure[]{\centering\includegraphics[width=0.48\textwidth]{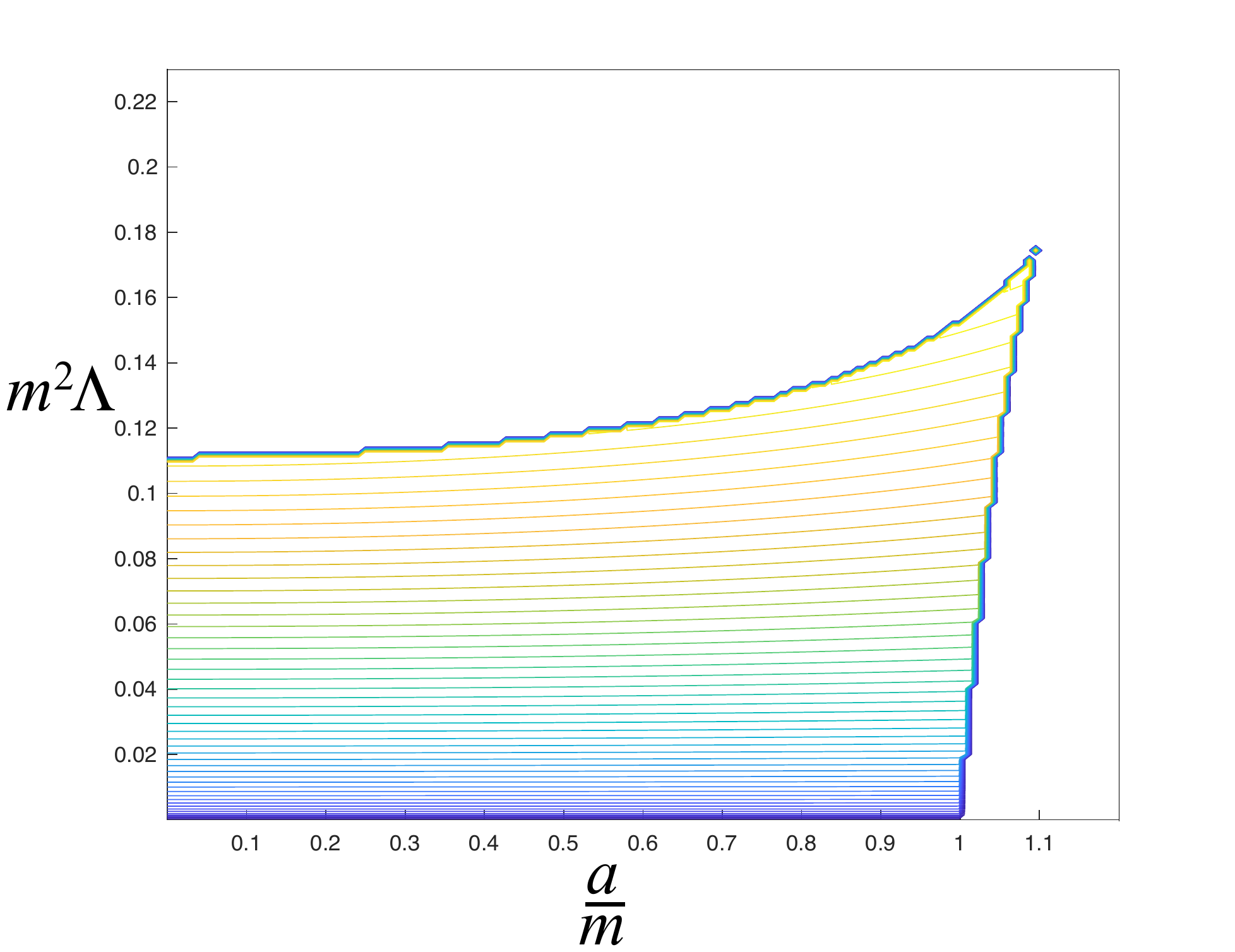}
\label{fig:constmass--Kerr-dS}}
\subfigure[]{\centering\includegraphics[width=0.48\textwidth]{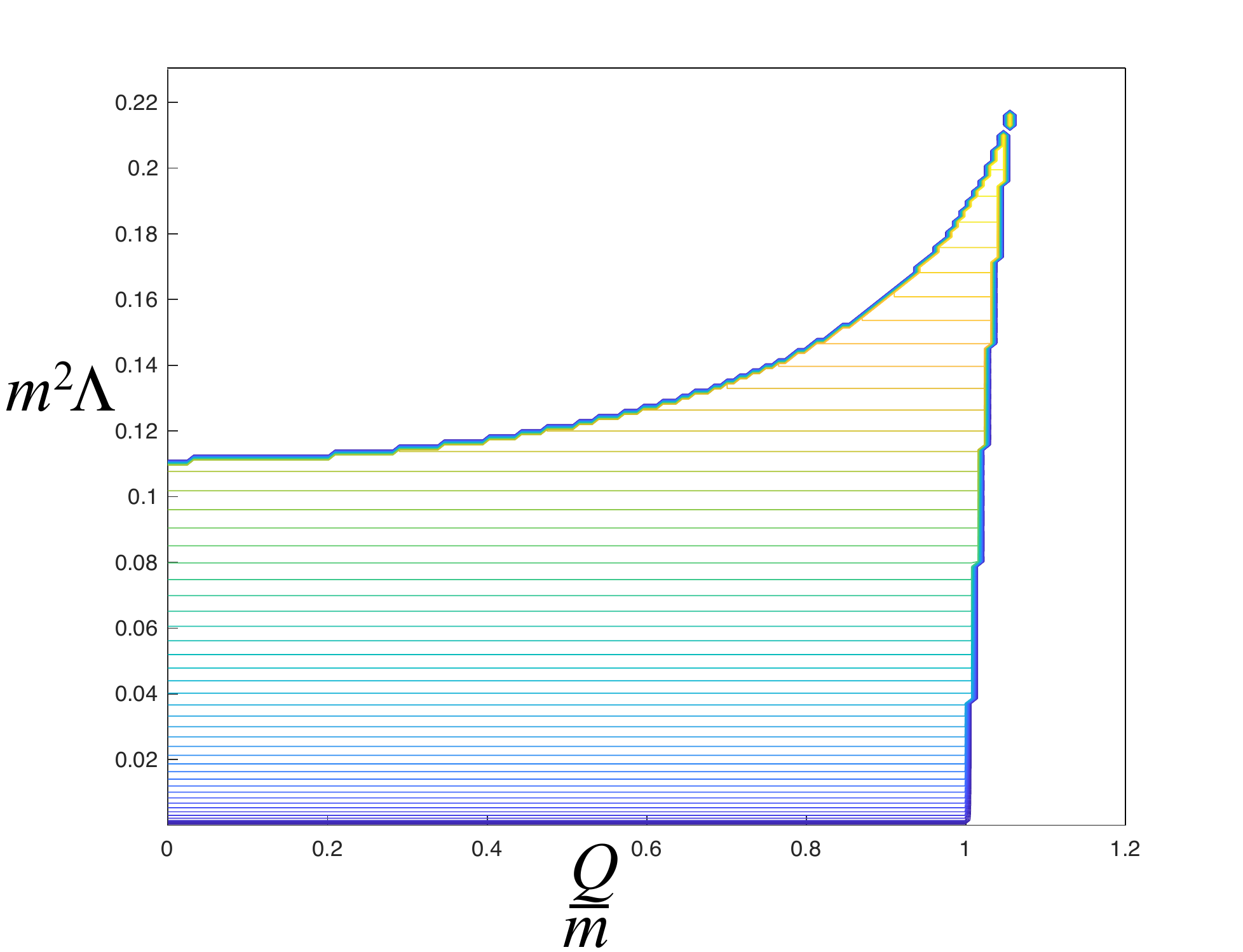}
\label{fig:constmass--RN-dS}}
\caption{\label{fig:constmass}(a)  Lines of constant physical mass $M$ in Kerr--de Sitter.  (b)  Lines of constant physical mass $M$ in Reissner--Nordstr\"om--de Sitter, which also are lines of constant mass parameter $m$.}
\end{figure}

Consider the case of $dJ<0$, with an increase in the black hole entropy $S_b$, corresponding to positive heat inflow, $Q_H$. Recall that $dJ<0$ corresponds to the system doing positive work, $W$, 
 There are three interesting categories of such trajectories:

\begin{itemize}

\item One such $dJ<0$ class of paths has the physical black hole mass $M$ decreasing, with $dS_b>0$.  This is precisely what  classic Penrose processes\cite{Penrose:1969pc,Penrose:1971uk,Denardo:1973pyo} (and superradiant modes) do\footnote{Superradiant modes  reduce $J$ and $M$ if they have azimuthal quantum number $m$ (not to be confused with the mass parameter of earlier sections) and a frequency $\omega$ such that~\cite{TACHIZAWA1993325,Gibbons:1977mu} $m\Omega_c<\omega <m\Omega_b$.}. For such processes, however, the area of the cosmological horizon also increases\footnote{This was found  by numerically exploring the parameter space, but in fact there is a proof  in ref.~\cite{Bhattacharya:2017scw}.}, $dS_c>0$. This means $Q_C>0$ and $Q_H>0$, but the system's energy changes since $dM<0$. So Penrose processes (some samples are shown in figure~\ref{fig:trajectories-penrose-Kerr-dS}) are not part of the operation of the continuous heat engine.

\item Another $dJ<0$ class of trajectories   has $dM>0$ with $dS_b>0$, but crucially they have $dS_c<0$, so they are not Penrose processes, since mass--energy is not taken from the black hole\footnote{In a true Penrose process or superradiant scattering event, the final outgoing component goes off to infinity. In de Sitter, scattering to infinity results in an encounter with the cosmological horizon, which is one reason that its area  increases in such processes. Perhaps these non--Penrose processes can be thought of as encounters for which the outgoing components do not escape far away enough from the black hole, so there is no reduction of $M$, and they do not cross the cosmological horizon so its area, and hence $S_c$, does not increase.}. For these, $W$~and~$Q_H$  are positive, but ~$Q_C<0$, meaning heat also flows in from the cold bath.  Some examples of such paths are shown in figure~\ref{fig:trajectories-non-penrose-Kerr-dS}. Since they change $M$, however, they do not represent continuous engine operation.

\item The $dJ<0$ class of trajectories that has $dM=0$  is the set that preserves the internal energy of the system while the heat flows take place, and while positive work is being done. This class will define the continuous engine for a general de Sitter black hole ({\it i.e.,} including additional work terms, coupling to other fields, and so forth.) In general, these paths display the classic heat engine operation, converting heat into work with some possible exhaust output (although as will become clear below for Kerr--dS and RN--dS, the exhaust is zero). Some examples of these paths are shown in figure~\ref{fig:constmass} for Kerr--dS and RN--dS.
\end{itemize}
 
 \noindent
 The general heat engine is subject to constraints from the second law of thermodynamics, to be discussed next.


\subsection{A Bound on Efficiency}
\label{sec:bound}
In the prototype continuous heat engine at the beginning of this section,  a key observation was made about the efficiency of the system~\cite{PhysRevLett.2.262}. As there are only three  energies involved, it is simply written as $\eta=(\omega_H-\omega_C)/\omega_H = 1-\omega_C/\omega_H$. There's a bound on it, which follows from the condition for population inversion, $n_2/n_1\ge1$. This is, using the Boltzman expressions~(\ref{eq:boltzmann}):
\begin{equation}
\exp\left\{\frac{\omega_C}{T_C}-\frac{\omega_H}{T_H}\right\}=\exp\left\{\frac{\omega_H}{T_C}\left(\frac{\omega_C}{\omega_H}-\frac{T_C}{T_H}\right)\right\}=\exp\left\{\frac{\omega_H}{T_C}(\eta_{\rm C}^{\phantom{C}}-\eta)\right\}>1 \ ,
\end{equation}
which amounts to the Carnot bound,  equivalent to the second law of thermodynamics:
\begin{equation}
\eta\le\eta_{\rm C}^{\phantom{C}}\equiv 1- \frac{T_C}{T_H}\ .
\end{equation}
More general continuous heat engines have been discussed widely in the literature since 1959 (particularly in the context of quantum thermodynamics and quantum open systems~\cite{doi:10.1080/00107514.2016.1201896,Halpern:2018bul}), and bounds on their efficiency are formulated directly from the second law  (see {\it e.g.,} ref.~\cite{2014ARPC...65..365K}).

For a de Sitter black hole moving along a continuous heat engine trajectory, it is natural to suggest that the efficiency must be bounded by Carnot as well, suggesting an important constraint from the second law of thermodynamics on the evolution of a de Sitter black hole, at least for the $dM=0$ paths. It should follow from the fact that the second law requires that the total entropy change must be zero or positive: $dS_a+dS_b\ge0$. In general, the most straightforward suggestion would be a bound at any point along a trajectory in parameter space. In terms of heat (for the case left of the lukewarm region where $T_b>T_c$):
\begin{equation}
\label{eq:efficiency}
\eta = \frac{W}{Q_H}= 1-\frac{Q_C}{Q_H} =1-\frac{T_cdS_c}{T_bdS_b}\leq 1 - \frac{T_c}{T_b}\ ,
\end{equation}
assuming $Q_C>0$ and $Q_H>0$. For a reciprocating engine, the analogue of this has an integration over a path. In reversible cases,  the temperature factors out  and the relation is saturated since the system state variables  go in a loop. For a continuous engine, it is more subtle. Conservation of energy allows for $0\leq dS_c/dS_b\leq T_b/T_c$, and the more strict, Carnot bound tightens the lower limit such that $1\leq dS_c/dS_b\leq T_b/T_c$.
This is a tight restriction on $dS_c/dS_b$ along a $dM=0$ path for a general de Sitter black hole, and is worth testing (with a caveat to be discussed below).

In the above it was assumed that  $dS_c/dS_b>0$. There is also the possibility that this quantity is negative. This is in fact what happens here, for the Kerr--dS and RN--dS examples. In both cases, along the~$dM=0$ paths, it was observed that $dS_c<0$ while $dS_b>0$, although their magnitudes are not equal in general. In the  conventions established here, $Q_C<0$  means that heat flows in {\it from}  the cold reservoir as well.   This does {\it not} mean that the efficiency of the engine is greater than unity.  In such cases, the efficiency of equation~(\ref{eq:efficiency}) is not the correct definition. Efficiency should include all the heat that flows into the system, and that is $Q_H+Q_C$ here, and the efficiency is therefore unity\footnote{It is tempting to interpret the positive inflow of heat from the cold bath in terms of $T_c$ there being a negative temperature, which would also fit (loosely) with the  observation of earlier sections  that there is a maximum energy level. This is commonly  discussed in the literature (see {\it e.g.,}~refs.~\cite{PhysRev.156.343,2016arXiv160605244C,2017CoTPh..68..347X}), but will not be explored here.}. In other words, the engines satisfy a bound, but only the trivial one following from the first law (energy conservation). It is important to note that since this is not a reciprocating engine (where after a cycle the system returns to its original state), conversion of all input heat into work is not in contradiction with the second law of thermodynamics. The simple case of an isothermal expansion of an ideal gas is another such example.

More general de Sitter examples can presumably have positive work paths with  $dM=0$  such that $dS_b>0$ but with $dS_c>0$. It would be interesting to find such examples, since as discussed above, the point--wise bound above constrains $dS_c/dS_b$. Equation~(\ref{eq:efficiency}) amounts to  a derivative condition:
\begin{equation}
\label{eq:bound-strong-Kerr-dS}
\eta=(\Omega_b+\Omega_c-2\Omega_\infty)\frac{1}{T_b}\left| \frac{dJ}{dS_b}\right| \le1-\frac{T_c}{T_b}\ ,
\end{equation}
(with the obvious exchange of $\Omega$ for potential $\Phi$  for charged case).

An escape clause to the above point--wise formulation of the bound is the fact that in general thermodynamics, heat flow should not always be written in the form $TdS$. The possibility of irreversibility (and $\eta$ falling  below a Carnot bound) is more easily accommodated when heat comes from integrating $TdS$ over a finite range, as happens in the gravity context in the reciprocating heat engines of ref.~\cite{Johnson:2014yja}. So it is possible that the bound above on de Sitter gravity solutions  is too strong. A weaker form of a bound might be to constrain  the heat and work over a whole trajectory,  writing instead:
\begin{equation}
W^{\rm tot}=-\int(\Omega_b+\Omega_c-2\Omega_\infty)dJ\ ,\quad {\rm and}\quad  Q^{\rm tot}_H=\int T_bdS_b\ ,
\end{equation} where the integrals are performed along the $dM=0$ path in parameter space, the weaker  bound is:
\begin{equation}
\label{eq:bound-weak-Kerr-dS}
\eta= \frac{W^{\rm tot}}{Q^{\rm tot}_H} \le1-\frac{T_C}{T_H}\ ,
\end{equation}
where now $T_H$ and $T_C$ are the highest and lowest temperatures that were encountered along the path. This weaker bound is easily satisfied for trajectories that start near the right hand or bottom borders of parameter space (see figure~\ref{fig:parameter-space}), since the bound tends to unity  for those, given that $T_C$ vanishes there. It is therefore perhaps less useful.

This is an important construction to explore further, a companion to the reciprocating ``holographic'' heat engines proposed in ref.~\cite{Johnson:2014yja}. Other de Sitter black hole solutions (perhaps coupled to other fields, matter, {\it etc})  can be studied in this light, and the $dM=0$ trajectories would be constrained by the bounds above. This could even be useful as a means of helping test the consistency of  more complex de Sitter solutions, as is now done with holographic heat engines for anti--de Sitter solutions (see {\it e.g.,} refs.~\cite{Mo:2017nes,Ahmed:2019yci,Zhang:2016wek}). The bound could   be used as a means of  ruling out (on the grounds of internal consistency of the gravitational thermodynamics)  trajectories of evolution in parameter space for de Sitter black holes. More generally, the idea of de Sitter black holes having a natural maser action due to their interaction with the heat baths might have applications in a cosmological and astrophysics setting, particularly in the early universe, so this is worth exploring further.

Two obvious supplements to the work explicitly described were explored, and are readily constructed. First, there exist heat engine trajectories to the right of the lukewarm region also. Since $T_c>T_b$, it is the cosmological horizon that connects to the high temperature reservoir, and the black hole  to the low. The bounds analogous to~equations~(\ref{eq:efficiency}, \ref{eq:bound-strong-Kerr-dS}, \ref{eq:bound-weak-Kerr-dS}) for such paths can be written as before, with the straightforward exchange of the labels $b$ and $c$. Second, refrigerators and heat pumps can be made by running the trajectories in reverse. The bounds on efficiency/performance of those machines are straightforward to write down. 

\section{Discussion}
\label{sec:discussion}

A key task of this project was to establish that the kinds of Schottky--like peaks seen previously~\cite{Johnson:2019vqf}  in extended black hole thermodynamics can be naturally present for  traditional black hole thermodynamics as well. A natural setting seems to be  de Sitter black holes, and the cases of  Kerr--de Sitter and Reissner--Nordstr\"om de Sitter were explored and exhibited. This is a companion to work presented recently for Schwarzschild--de Sitter~\cite{Dinsmore:2019elr}, which is a special case of the K--dS and RN--dS systems. A point emphasized in refs.~\cite{Johnson:2019vqf,Johnson:2019olt} is that the peaks signal a striking (and perhaps unexpected) fact about the underlying degrees of freedom of the system: There is a finite window of available energies that the degrees of freedom can occupy. 

The energy window can be a few discrete levels in a spin or other similar system, or it can be a cutoff on some larger more complex spectrum, as noted in ref.~\cite{Johnson:2019olt}, and reviewed here. The intricate physics already known for de Sitter black holes perhaps makes the latter more likely, and the subsequent thermodynamic interpretation presented assumes this (perhaps in polite disagreement with some of the details of the approach of ref.~\cite{Dinsmore:2019elr}).

The fact that de Sitter's structure suggests that some features can be described using equilibrium thermodynamics quantities, while at the same time it is manifestly not an equilibrium system (having two temperatures, for example) is a well--known puzzle. It was pointed out, however that  there are known systems for which this is not an obstruction to interpreting the physics. The laser and maser boast such features, with two (at least) distinct effective temperatures needed to achieve  population inversion, a non--equilibrium state. The two key features they need are a maximum energy, and a channel for outputting work.  

Kerr--de Sitter and Reissner--Nordstr\"om--de Sitter have precisely these same features and it was argued that there exist clear analogues of the maser action for these and even more general de Sitter black hole systems. They can operate as continuous heat engines (another way of describing masers), corresponding to a mass--preserving trajectory in parameter space. The output channel is a thermally--driven reduction of angular momentum (or charge, or other parameters associated with work terms). They are a complementary idea to the reciprocating black hole heat engines  proposed in ref.~\cite{Johnson:2014yja} in the context of extended thermodynamics. Note that they should not be confused with the Blandford--Znajek engines~\cite{Blandford:1977ds} that probably drive active galactic nuclei, which are not heat engines, nor should they be confused with the recently proposed astrophysical laser mechanisms proposed to be powered by superradiance by black holes (BLASTs~\cite{Rosa:2017ury}). There, the (axion) environment is doing the lasing/masing. Here, it is the black hole plus de Sitter system itself that is the maser/laser. However, it is not hard to imagine a role for them in cosmological/astrophysical settings where significant positive $\Lambda$ is relevant. Perhaps the early universe could be such a setting.

An immediate output of this continuous heat engine construction is a suggested bound (due to the second law of thermodynamics) on how much of this (thermally driven) angular momentum reduction can occur along trajectories along the parameter space. Whether the continuous heat engine or maser/laser language is attractive or not, the point is that {\it the second law of thermodynamics should place constraints on how the de Sitter black hole system classically explores its parameter space}, and in this language it is simply a bound on the engine efficiency.    It should be interesting to explore this paper's constructions for other de Sitter black holes beyond K--dS and~RN--dS. 

Versions of this construction should be quite natural in extended thermodynamics, where the cosmological constant is treated as a dynamical variable. After all, the original black hole heat engine construction of ref.~\cite{Johnson:2014yja} was motivated by the opportunity to do mechanical work with the~$pdV$ terms available in the first law, as formulated in ref.~\cite{Kastor:2009wy}, so continuous counterparts for them would be natural. Presumably, it would be the $dU=0$ trajectories in (extended) parameter space that would be the continuous engine trajectories, leaving variable the mass $M$ of the black hole, since it is the enthalpy in that setting: $M=U+pV$. Work would now involve changing the volume between the horizons~\cite{Dolan:2013ft}, and the fact that the pressure is negative could have a natural interpretation in terms of the heat engine work channel. 

Finally, as remarked at the end of the introduction (section~\ref{sec:introduction}), the idea that there is a description of (positive~$\Lambda$) quantum spacetime physics that involves a limit on the energy that can be carried by the microscopic degrees of freedom, as suggested by the Schottky--like peaks, is very much worth exploring. It may well have wider consequences that can be put to use.

 
\section*{Acknowledgements}
The work of CVJ  was funded by the US Department of Energy  under grant DE-SC 0011687.  Some of this work (preparing a revised manuscript) was done at the Aspen Center for Physics, which is supported by National Science Foundation grant PHY-1607611. Helpful questions and comments from Ibrahima Bah, David Kastor, Mitchell Porter, Felipe Rosso, Eva Silverstein, Lenny Susskind, Leo Stein, and Jennie Traschen are gratefully acknowledged.  CVJ thanks, as always, Amelia for her support and patience.

\bibliographystyle{utphys}
\bibliography{de_sitter_black_holes}

\end{document}